\documentclass[12pt]{aa}
\usepackage[utf8]{inputenc}
\usepackage{graphicx}
\usepackage{natbib}
\usepackage{amsmath}
\usepackage{booktabs}
\usepackage{pdfpages}
\usepackage{listings}
\usepackage{txfonts}
\usepackage[backref,breaklinks,colorlinks,citecolor=blue]{hyperref}

\newcommand{\myimage}[1]{\begin{center}\includegraphics[angle=0, width=0.47\textwidth]{images/#1}\end{center}}

\newcommand{\myimagelarge}[1]{\begin{center}\includegraphics[angle=0, width=0.97\textwidth]{images/#1}\end{center}}

\DeclareMathSymbol{\mlq}{\mathord}{operators}{``}
\DeclareMathSymbol{\mrq}{\mathord}{operators}{`'}
\DeclareMathOperator{\erf}{erf}

\newcommand{\cavg}{\langle\mathbf{c}(z)\rangle}
\newcommand{\covariance}{\mathbf{C}_{int}(z)}
\newcommand{\rnfw}{r_{\rm NFW}}

\newcommand{\XMatch}{SDSS-UKIDSS-AllWISE~}

\begin{document}

\title{The Integrated Cluster Finder for the ARCHES project}
\date{\today}
\author{Alexey~Mints\inst{\ref{inst_AIP},\ref{inst_MPS}}\thanks{email: mints@mps.mpg.de} \and Axel~Schwope\inst{\ref{inst_AIP}} \and Simon~Rosen\inst{\ref{inst_L}}  \and François-Xavier~Pineau\inst{\ref{inst_F}}  \and Francisco~Carrera\inst{\ref{inst_S}}}
\institute{Leibniz-Institut für Astrophysik Potsdam (AIP), An der Sternwarte 16, 14482 Potsdam, Germany \label{inst_AIP}
\and
Max Planck Institute for Solar System Research, Justus-von-Liebig-Weg 3, 37077 Göttingen, Germany \label{inst_MPS} 
\and
The University of Leicester, University Road, Leicester, LE1 7RH, United Kingdom\label{inst_L}
\and
Observatoire Astronomique de Strasbourg, 11 rue de l'Université, 67000 Strasbourg, France\label{inst_F}
\and
Instituto de Fisica de Cantabria (CSIC-UC), Avda. de los Castros s/n, 39005 Santander, Spain\label{inst_S}
}

\abstract{Clusters of galaxies are important for cosmology and astrophysics. They may be discovered through either the summed optical/IR radiation originating from their member galaxies or via X-ray emission originating from the hot intracluster medium. X-ray samples are not affected by projection effects but a redshift determination typically needs optical and infrared follow-up to then infer X-ray temperatures and luminosities.}
{We want to confirm serendipitously discovered X-ray emitting cluster candidates and measure their cosmological redshift through the analysis and exploration of multi-wavelength photometric catalogues.} 
{We developed a tool, the Integrated Cluster Finder (ICF), to search for clusters by determining overdensities of potential member galaxies in optical and infrared catalogues. Based on a spectroscopic meta-catalogue we calibrated colour-redshift relations that combine optical (SDSS) and IR data (UKIDSS, WISE). The tool is used to quantify the overdensity of galaxies against the background via a modified redMaPPer technique and to quantify the confidence of a cluster detection.}
{Cluster finding results are compared to reference catalogues found in the literature. The results agree to within 95-98\%. The tool is used to confirm 488 out of 830 cluster candidates drawn from 3XMMe in the footprint of the SDSS and CFHT catalogues.}
{The ICF is a flexible and highly efficient tool to search for galaxy clusters in multiple catalogues and is freely available to the community. It may be used to identify the cluster content in future X-ray catalogues from XMM-Newton and eventually from eROSITA.}

\maketitle

\section{Introduction}
Clusters as building blocks of the Universe are important objects for astrophysics and cosmology and large-scale surveys across the electromagnetic spectrum have been  performed to detect and characterize them in considerable number. The techniques used address different radiation properties of the constituents of a cluster.  

In X-rays clusters are seen through the thermal emission of the hot intracluster gas. X-ray observations may provide important information concerning cluster properties such as the temperature,  density, and mass.
For bright clusters the redshift can also be obtained directly from the X-ray data by measuring the wavelength of emission lines in the X-ray spectrum. For the more typical case with insufficient number of photons, one needs to complement X-ray observations with optical or infrared data, firstly, to confirm the nature of the object as a cluster and also to achieve a reliable redshift determination. Of order $\sim 2 \times 10^3$ clusters were discovered through their extended X-ray emission. Still most of the X-ray selected clusters are based on ROentgen SATellite (ROSAT) discoveries but XMM-Newton is ramping up thanks to its high sensitivity and despite its comparatively small survey area \citep[see e.g.~][]{clerc+14, mehrtens+12, Takey2, Takey3}.  

In the optical and at infrared wavelengths clusters are detected as overdensities of their member galaxies that are orbiting in their dark matter halo. 
A large suite of techniques were developed to find those overdensities
including Voronoi tessellations~\citep{Gerke}, matched filter algorithms \citep{postman}, and a variety of red sequence techniques \cite[for examples see][and references therein]{Koester2007, Wen2009, Rykoff2013}. 
The red sequence method requires good photometry in several bands, preferably enclosing the $4000$\,\AA\ break. This limits the applicability of such a cluster finder when only optical bandpasses are used. However, owing to the unprecedented combination of imaging depth and solid angle, some 50000 to 100000 clusters were detected in the Sloan Digital Sky Survey (SDSS) \citep[see][]{Rykoff2013, Wen2009}. Cluster searches based on SDSS are limited to $z \lesssim 0.55$. The method can in principle be improved for higher redshifts  if infrared data are included. This is one of the achievements presented in this paper. The Integrated Cluster Finder (henceforth ICF) searches multiple catalogues for overdensities of passive red galaxies in the vicinity of extended X-ray sources. If successful, the ICF confirms X-ray detection is  due to a massive dark matter halo with hot gas and galaxies  trapped in it, and it measures the cosmological redshift.

In the context of this paper a cluster candidate is regarded as confirmed or a cluster as identified if its cosmological redshift could be measured through a set of concordant redshifts of its likely member galaxies.

At millimeter wavelengths clusters can be detected out to high redshifts via Sunyaev-Zel'dovich (SZ) effect, which is caused by the inverse Compton scattering of cosmic microwave background (CMB) photons with electrons in the hot intra-cluster gas. Just as for X-ray searches SZ clusters are detected as a single objects. An important property of the SZ effect is that the surface brightness of the thermal SZ effect is independent of redshift. The Sunyaev-Zel'dovich surveys became mature only in recent years, and the first Planck-discovered catalogue lists some 1227 entries \citep{SPTSZ, PlanckSZ}.

The ICF was developed as part of the Astronomical Resource Cross-matching for High-Energy Studies (ARCHES), which is an EU-FP7 funded small-scale research project; ARCHES aims to develop tools for the correlation of X-ray sources detected by XMM-Newton with multi-wavelength data resources. Two main cases need to be considered, correlations for non-resolved (point-like) and for resolved (extended) X-ray sources. While the former class of sources are primarily active galactic nuclei (AGN) with a growing fraction of stellar objects towards the galactic plane, the latter are clusters of galaxies in the first place. 

The basis of the ARCHES project in general and of this subproject in particular is 3XMM, the catalogue of all X-ray sources detected by XMM-Newton over the past 14 years \citep{3XMMDR5}. For each detected X-ray source, many parameters are derived and listed in 3XMM, and two of those are related to the extent of the X-ray source: a likelihood that the profile of the source exceeds the point-spread profile of the X-ray mirror assembly, {\it ext\_ml}, and the measured extent of an assumed $\beta$ profile fitted to the sourc. The $\beta$ profile has the form $S(r) = S(0)[1 + (r/r_c)^2]^{-3\beta + 1/2}$, where $S(0)$, $\beta$ and $r_c$ are the profile parameters. Mainly these two parameters were used to prepare an input catalogue for cluster identification in this project.

Finding and correlating multi-wavelength counterparts to X-ray point and extended sources is fundamentally different. For X-ray point sources one typically seeks one matching counterpart per external catalogue at a given celestial position. Here positional uncertainties play a very important role. This matching and the corresponding tool will be presented by other participants of the ARCHES  \citep[see][and ARCHES web page\footnote{\url{http://www.arches-fp7.eu/index.php/tools-data/online-tools/cross-match-service}}]{Pineau2015, Pineau2016}. For clusters of galaxies, the X-ray object, hot intergalactic medium, and potential optical or infrared counterparts, member galaxies are completely different entities, and therefore the correlation is more involved. We used the novel ARCHES cross-matching tool in the ICF to obtain magnitudes and photometric redshifts of potential cluster member galaxies from a variety of sources.

A database containing parameters for all X-ray sources and their correlated entries in other catalogues was developed within ARCHES. For confirmed clusters, the likely redshift, a finding chart, a plot of the multiplicity function, and the list of likely members galaxies are stored in the database and made available to the community.
The ICF does not attempt to include the SZ signal in the detection process. 

The paper is organized as follows: In section~\ref{sec:general} we provide a general description of the Integrated Cluster Finder algorithm, in section~\ref{sec:implementation} we present our implementation of the algorithm, in section~\ref{sec:tests} we report on tests with known clusters, in section~\ref{sec:fom} we predict completeness and false positives fraction levels of our cluster catalogue. The construction of the catalogue is described in section~\ref{sec:icfcatalogue}. In section~\ref{sec:discussion} we provide a summary of the results and an outlook.

\section{General description}\label{sec:general}
The Integrated Cluster Finder (ICF) implements a red MaPPer-like~\citep[see][]{Rykoff2013} algorithm to search for clusters in the space spanned by the positions and the colours of their member galaxies. A first major difference with respect to the published redMaPPer method is that prior information concerning the position of the cluster from the detected X-ray position is integrated in our method. A further major difference is the inclusion of near- and mid-infrared photometric data whose cross-colours need to be calibrated by a large spectroscopic training set.

In general, the method is based on the fact that luminous red galaxies (LRGs) have (on the average) redshift-dependent colours. One can thus estimate the redshift of the member galaxies from a calibrated redshift-colour relation and the redshift of the cluster via averaging over all potential member galaxies.

Following Rykoff et al.~a multiplicity function $\lambda(z)$ is introduced, which is a measure of the background-corrected number of galaxies in the cluster. This function is calculated on a pre-defined grid of redshifts. Background density of LRGs, their colours, and the colour covariance matrix have to be determined for each redshift value on the grid.

The only input the algorithm takes is a position on the sky, which is the expected cluster position. This is driven by the primary use of this cluster finder, namely the search for, and redshift determination of, X-ray selected galaxy clusters. 

The algorithm contains four main steps:
\begin{enumerate}
  \item Calibration of the redshift-colour relation: For a chosen set of external catalogues one obtains average colours and colour-colour covariance matrices of red galaxies for each redshift. An alternative version of the ICF that uses pre-fabricated values of photometric redshifts (like those provided with SDSS, Canada-France-Hawaii Telescope
Legacy Survey (CFHTLS), or Advance Large Homogeneous Area Medium Band Redshift Astronomical (ALHAMBRA) catalogues) was also developed.
  \item Determination of the background density to be incorporated into the determination of a multiplicity at the cluster position.
  \item Loading and filtering the data on galaxies.
  \item Cluster search: Identification of possible cluster member galaxies and measurement of the cluster redshift.
\end{enumerate}

Steps 1 and 2 need to be carried out only once for each of the optical and infrared catalogues, while steps 3 and 4 need to be performed for every cluster candidate individually.

We acquired data from various optical and infrared catalogues in a circular region of radius $R = 8'$ around the position of each extended 3XMM source (for a more detailed description of the kind of downloaded data see below). This is a compromise between completeness and detection efficiency and was chosen to limit the size of data to be downloaded and inspected for each position. Owing to this choice, our approach is inefficient for the detection of nearby clusters because  clusters appear larger than $8'$ on the sky at low redshift. The chosen radius corresponds to 1 Mpc co-moving radius at redshift $z=0.12$. However, the restriction to $8'$ does not seem to be severe since most nearby clusters are already known and have their redshifts measured. In addition, bright, nearby (hence very extended) clusters are not reliably detected as single entities by the XMM-Newton source detection chain and are thus partially screened away by the filters applied to generate 3XMM.

\section{Implementation}\label{sec:implementation}
We describe step 4 first, as it does not depend on the catalogue chosen. We proceed with the description of steps 1 to 3 as they were implemented in the current version of the ICF.

\subsection{The multiplicity function $\lambda(z)$}\label{sec:lambda}

A multiplicity function $\lambda$ as function of the redshift $z$ is defined in Eq.~\ref{eq:lambda} as sum of cluster membership probabilities for all galaxies in the field of view. This function is calculated as the solution of

\begin{equation}\label{eq:lambda}
  \lambda(z) = \sum_{r_{\rm ang} < \tilde{R}(z)} \frac{\lambda(z) u(z, x)}{\lambda(z) u(z, x) + b(z, x)},
\end{equation} 
where $r_{ang}$ is the angular separation between the galaxy and the given position of the extended X-ray source, and $\tilde{R}(z)$ is the minimum between $R_s$, the distance corresponding to 1 Mpc projected distance at redshift $z$, and $R = 8'$.

In this equation the following definitions and quantities are used:
\begin{eqnarray}
  x &=& (r_{ang}, m, \chi^2), \\
  u(z, x) &=& \Sigma(r_{ang}) \theta(m) p_\nu(\chi^2), \label{eq:u} \\
    p_\nu(\chi^2) &=& \frac{\gamma(\nu/2, \chi^2/2)}{\Gamma(\nu/2)}, \label{eq:chi2}\\
 \theta(m) &=& \exp\left({-10^{-0.4(m - m_*)}}\right), \label{eq:luminosity} \\
 \Sigma(r_{ang}) &=& \frac{w}{(r_{ang}/R_s)^2 - 1}f(r_{ang}/R_s), \label{eq:nfw} \\
 f(t) &=& \left\{ 
  \begin{array}{l l}
    1 - \frac{2}{\sqrt{t^2-1}} \tan^{-1} \sqrt{\frac{t-1}{t+1}}, & t > 1 \\
    0, & t = 1 \\
    1 - \frac{2}{\sqrt{1 - t^2}} \tanh^{-1} \sqrt{\frac{1-t}{t+1}}, & t < 1
  \end{array}\right.
,\end{eqnarray}
where $m$ is the main magnitude of the candidate member galaxy (for a definition of 'main' magnitude see below) and $m_*$ is the magnitude of a galaxy with a mass of $2\times 10^{11} M_{\odot}$.

The function $u(z, x)$ is the normalized cluster profile. It is the product of a surface density $\Sigma(r_{ang})$, a luminosity function $\theta$, and a membership probability $p_\nu(\chi^2)$. A Schechter function with $\alpha = -1$ was used for the luminosity function $\theta(m)$
. The function $\Sigma(r_{ang})$ was described with a NFW universal density profile \citep[NFW: Navarro-Frenk-White;][]{NFW}. Both $\Sigma(r_{ang})$ and $\theta(m)$ are normalized so that their integrals taken over all possible values of $r_{ang}$ or $m$ is unity. 

The quantity $p_\nu(\chi^2)$ gives the probability for a given galaxy to be at a given redshift and is calculated as the inverse of the cumulative distribution function (CDF) of the $\chi^2$ distribution, where $\gamma$ is the incomplete gamma function and $\nu$ the number of degrees of freedom (equal to the number of colours used). More details on the $p_\nu(\chi^2)$ definition are given in Appendix~\ref{sec:pnu}.

A NFW surface density profile was used here with a normalization $w$ chosen to satisfy the equation $\int_0^\infty \Sigma(r) 2 \pi r dr = 1$. Equation~\ref{eq:nfw} cannot be calculated numerically for $r = R_s$, so within $r = R_s\times (1 \pm 0.01)$ it was approximated by a simple linear function. Also, following \cite{Rykoff2013}, a cut-off is imposed to prevent overweighting of the central parts of the cluster, $\Sigma(r) = \Sigma(r_0), \textrm{if}\ r < r_0 = 0.15\, \textrm{Mpc}$.
This allows for an optical-to-X-ray offset, which is typically of the order of 50-100 kpc \cite[see, for example,][]{Takey2} and does not allow a single central object to dominate the cluster. The  offset might appear because of an unrelaxed state of the cluster or errors in the positioning of the X-ray emission peak.

The NFW profile used here (or any other symmetric profile) biases our finding approach towards relaxed symmetric clusters. The strength of this bias is yet to be studied.

In Eq.~\ref{eq:lambda} $b$ is the background density, which is a tabulated function of redshift, magnitude, and  $p_\nu(\chi^2)$, which is derived below in Section~\ref{sec:background}.

Two simple critical cases might help to understand the nature of the $\lambda(z)$ function. If $b = 0$ then $\lambda = N$ -- simply the number of galaxies in the field of view. If $b = const$ and $u = const$ (i.e. all galaxies are exactly the same) then $\lambda = N - b/u$. For further properties of the multiplicity function, see \cite{Rykoff2013}.

The solution of Eq.~\ref{eq:lambda} is obtained iteratively for each value of $z$. The procedure gives cluster membership probabilities for all galaxies in the field as 
\begin{equation}
  p_{\rm mem}(z, x) = \frac{\lambda(z) u(z, x)}{\lambda(z) u(z, x) + b(z, x)}.\label{eq:pmem}
\end{equation} 

We search for clusters of galaxies at the peaks of the $\lambda(z)$ function.

\subsection{Inclusion of spectroscopic data}\label{sec:specz}

Spectroscopic data were used whenever possible. To achieve this, the spectroscopic meta-catalogue (composed of SDSS-III's Baryon Oscillation Spectroscopic Survey (BOSS) and The VIMOS Public Extragalactic Redshift Survey (VIPERS) catalogues) was used as an extra catalogue for cross-matching. When calculating the function $u(z, x)$ in Eq.~\ref{eq:u} $p_\nu$ now depends on $z$, not on $\chi^2$, so the following equation is used in place of Eq.~\ref{eq:chi2} for objects with a spectroscopic counterpart with redshift $z_{sp}$:
\begin{equation}
  p_{\nu}(z) = \left\{
   \begin{array}{ll}
    1. & \textrm{if} |z - z_{sp}| < \Delta z \\
    \exp - \frac{0.5(z-z_{sp})^2}{\Delta z^2}& \textrm{if} |z - z_{sp}| \ge \Delta z  
   \end{array}\right.
   ,
\end{equation}
where $\Delta z$ equals the step-size of the used redshift grid. This value is chosen to ensure that every galaxy with known spectroscopic redshift contributes to at least two points
of the redshift grid. Thus galaxies with spectroscopic redshifts contribute to $\lambda(z)$ at an interval that is wider than a typical error of spectroscopic redshift. This accounts for the case when the $\lambda(z)$ peak is offset to a galaxy with spectroscopic redshift, for example, because of photometry uncertainties. If for a given detection there are one or more likely members with known spectroscopic redshift, then we report their average spectroscopic redshift together with the photometric cluster redshift estimate.

\subsection{Use of photometric redshifts}
For some catalogues used here, namely those from CFHT and ALHAMBRA, photometric redshifts are provided. For such a case, we modified the approach used in Section~\ref{sec:lambda} by replacing the Eq.~\ref{eq:chi2} with
\begin{eqnarray}
  p_{\nu}(z) &=& \frac{1}{\Sigma_{photoz} \sqrt{2 \pi \sigma_{photoz}^2}} \exp - \frac{(z-z_{phot})^2}{2 \sigma_{photoz}^2}, \\
  \sigma_{photoz}^2 &=& \delta z_{phot}^2 + 4 \Delta z^2, \\
  \Sigma_{photoz} &=& \erf\left(\Delta z \sqrt{\frac{2}{\sigma_{photoz}^2}}\right),
\end{eqnarray}
where $z_{phot}$ and $\delta z_{phot}$ are the value and error of the photometric redshift of a galaxy. An additional constraint is that $|z - z_{phot}| < 4 \Delta z$ should hold, i.e. the galaxy redshift should not be very different from the redshift under consideration. For galaxies outside of this range $p_{\nu}(z) = 0$. Therefore we added $\Sigma_{photoz}$ as a normalization factor, so the integral of $p_{\nu}(z)$ over all redshifts is one. We used a wider range than for spectroscopic redshifts, as photometric redshift errors are  substantially larger.

For the photometric redshifts the background function can be simplified. It is  defined on a grid of redshifts and magnitudes, that is
\begin{equation}
  b(z, m) = \sum_{M} p_{\nu}(z) / \Delta m,
\end{equation}
where $\Delta m$ is a grid step in magnitudes and the summation subset $M$ is defined as a set of galaxies for which their main magnitude $m_{main}$ falls into the corresponding bin in the magnitude grid and $p_{\nu}(z) > 0$.

If for a galaxy in the photometric redshift catalogue there is a known spectroscopic value, then the latter redshift is used together with its spectroscopic redshift error. 

\subsection{Peak finding and redshift determination}\label{sec:peakfinding}

The multiplicity function $\lambda(z)$ is calculated for each redshift value on a grid.  The grid in $z$ spans from 0.02 to 0.80 with a step-size of 0.01 for the combination of  SDSS, UKIRT Infrared Deep Sky Survey (UKIDSS) and AllWISE catalogues, and the upper limit was set to 1.4 (see secttion \ref{sec:icf_code}) for the photo-z catalogues. 
The step size was already chosen to be smaller than an average photometric redshift error with this kind of data over a wide redshift range \citep[see the review of photo-z methods in][]{Abdalla2011}. As the ICF cannot extract more redshift information than is normally provided by photo-z methods, an even smaller step size would not improve the cluster finding reliability and the redshift accuracy. 

Next one needs to find significant peaks and determine the most likely cluster redshift from $\lambda(z)$. Complexities may arise (1) as there might be more than one galaxy assembly along the line of sight and thus $\lambda$ might show more than one peak, and (2) the rather low precision of $\lambda$ measurements at higher redshifts due to large photometric errors. Two examples for the graph of the $\lambda(z)$ function are shown in figure~\ref{fig:lambda_example}: one with a single significant peak and a second with two significant peaks.

\begin{figure}[t]
  \includegraphics[width=0.47\textwidth, angle=0,clip=]{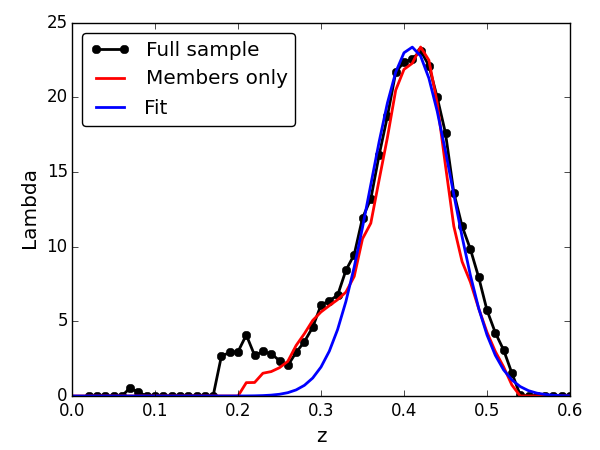}
  \includegraphics[width=0.47\textwidth, angle=0,clip=]{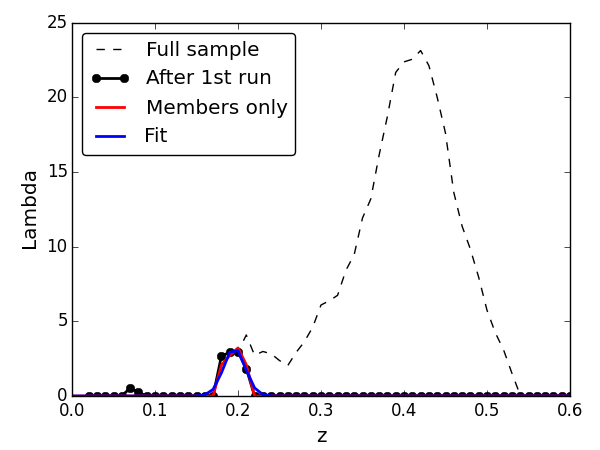}
  \caption{Peak finding algorithm illustration. Multiplicity function for the input sample is shown  as black curves. A subsample of possible members of a cluster is selected. Multiplicity function of this subsample (red curves) is then fitted with a Gaussian (blue curves). Possible members are then removed from the sample and the rest are taken as an input for the next iteration. \textit{Upper} panel shows the first iteration. \textit{Lower} panel shows the second iteration with the total multiplicity function shown with a thin dashed line for reference. See text for details.}\label{fig:peak_finding}
\end{figure}

Peaks in the $\lambda(z)$ function are determined in an iterative manner. 
The process is illustrated in Figure~\ref{fig:peak_finding}. The upper panel shows the first iteration: an initial candidate redshift $\tilde{z}_0$ is selected as the position of the absolute maximum of the $\lambda$ function in the whole interval covered by the redshift grid (black line): $max(\lambda(z)) = \lambda(\tilde{z}_0) = \lambda_0$. Then the individual membership probability $p_{mem}(\tilde{z}_0, x)$ for all galaxies within $r < \tilde{R}(z) = 1 \textrm{Mpc}$ at redshift $\tilde{z}_0$ is calculated. This initial list of probabilities is sorted in descending order. From the sorted list the first $n$ members are selected so that $\sum_{i=0}^n p_{mem}(\tilde{z}_0, x_i) \ge 0.9 \lambda_0$ and $\sum_{i=0}^{n-1} p_{mem}(\tilde{z}_0, x_i) < 0.9 \lambda_0$. Those selected galaxies give
approximately 90\% of the total multiplicity. The remaining 10\% is contributed by several galaxies with low membership probability. Then the cluster finder is applied again, this time using only the selected galaxies. This gives a new function $\lambda'(z),$ which needs to be divided by 0.9 since only 90\% of the initial multiplicity estimate was used. The revised function $\lambda'(z)$ (red line) is then fitted with a Gaussian function $\lambda_{fit} = \lambda_0 \exp{\frac{(z - z_0)^2}{2 (\Delta z_0)^2}}$ (blue line). This gives a most likely redshift $z_0$ for the current cluster candidate, which is close to $\tilde{z}_0$, a redshift error $\Delta z_0$, and a multiplicity value $\lambda_0$ . The selected $n$ galaxies are removed from the sample and the cluster finder is re-run. The resulting second iteration is illustrated in the bottom panel of Figure~\ref{fig:peak_finding}, where we have plotted $\lambda(z)$ for all galaxies with dashed line for reference (same data was shown with black line in the upper plot). This provides a new $\lambda(z)$ function with a new candidate redshift $\tilde{z}_1$ (solid black line). Then we again calculate $\lambda'(z)$ for members adding up 90\% of multiplicity (red line), which is then fitted with a Gaussian function (blue line). On both images the peak position, width, and height of Gaussian fits (blue lines) is nearly the same as for $\lambda'(z)$ data (red lines).

The procedure of galaxy elimination is repeated as long as $\lambda > 1$ for the new cluster candidate at redshift $\tilde{z}_n$. In the example we stop after the second iteration because the third peak (a very small bump at around $z = 0.08$ on Figure~\ref{fig:peak_finding}) has $\lambda < 1$.

The decision for a $90\%$ threshold is a compromise. If the threshold chosen was much higher, then galaxies not regarded as members tend to show up by building a secondary peak close to the main peak, thus producing many spurious cluster detections. If a  lower threshold was chosen, it is likely that separate structures were merged into one and, in this way, somehow produced artificially large errors for the redshift of the one detected cluster. We chose to select galaxies adding up 0.9 of the total $\lambda$ rather than selecting on membership probability $p_{mem}$ because the average value of $p_{mem}$ for cluster members decreases with redshift as the photometry becomes less and less precise. This method of member selection is similar to that used in the redMaPPer algorithm \citep[see][]{Rykoff2013}.

\begin{figure}[t]
  \includegraphics[width=0.47\textwidth, angle=0,clip=]{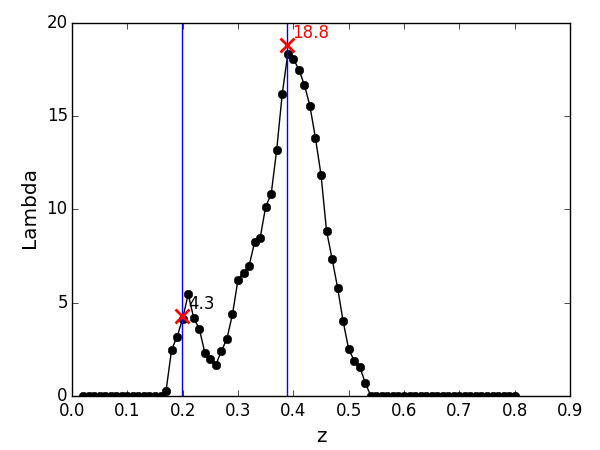}
  \includegraphics[width=0.47\textwidth, angle=0,clip=]{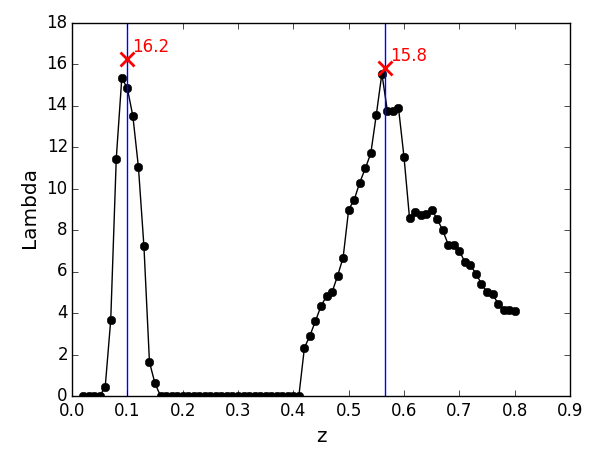}
  \caption{Examples of $\lambda(z)$ functions for two fields with one (top) and two (bottom) detections. Red crosses and vertical blue lines indicate detections. Values attached to crosses are for the measured $\lambda$, in red for non-spurious and in black for spurious detections (see Section~\ref{sec:fom}).}\label{fig:lambda_example}
\end{figure}

\begin{figure}[t]
  \includegraphics[width=0.47\textwidth, angle=0]{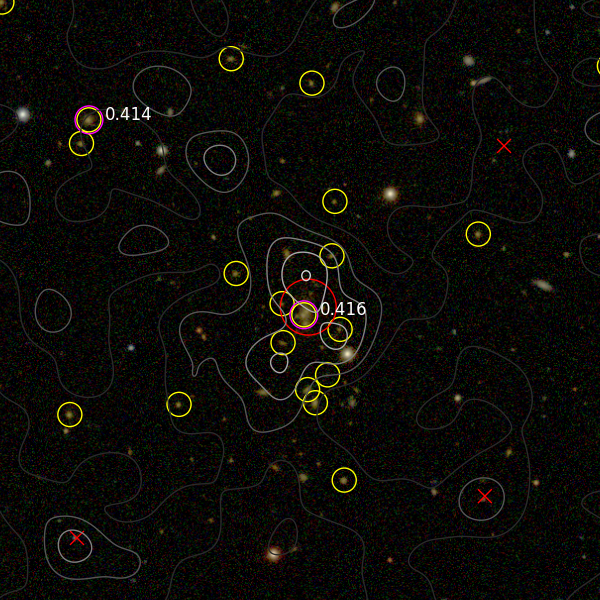}\vspace{1mm}
  \includegraphics[width=0.47\textwidth, angle=0]{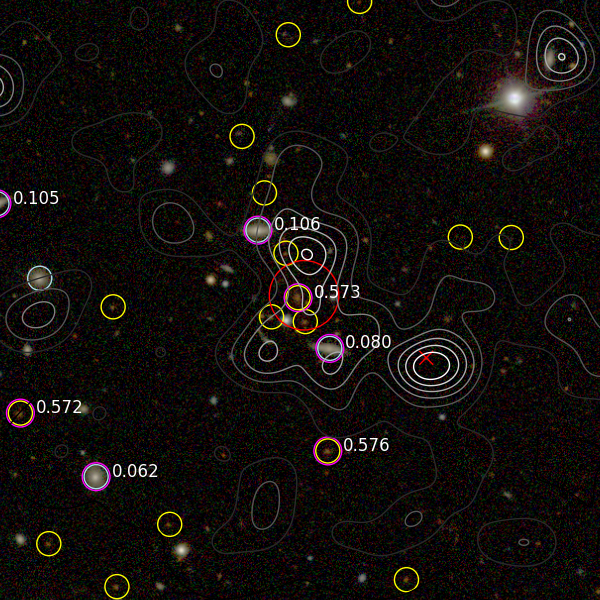}
  \caption{Finding charts for two cluster candidates from Figure~\ref{fig:lambda_example}. Coloured 2' by 2' SDSS images are overlaid with XMM X-ray contours, produced from co-added MOS and PN images in XMM bands 2 and 3. The red circle indicates the extended X-ray source position and extent as provided by 3XMM. The red crosses indicate the positions of point-like X-ray sources. The yellow and light blue circles denote likely cluster members with spectroscopic redshift indicated by purple circle and label where possible. See text for details.}\label{fig:charts}
\end{figure}

Two examples of peak finding are given in Fig.~\ref{fig:lambda_example}. Red crosses and vertical blue lines indicate possible cluster detections. Values attached to crosses are for the measured $\lambda$, in red for reliable and in black for spurious detections (for the definition of a spurious detection see section \ref{sec:fom}). In the upper panel there is only one reliable detection
at $z = 0.39 \pm 0.05$ with $\lambda = 18.8$. The secondary peak at $z=0.2$ was regarded as spurious. In the lower panel, a more complicated case with two detections is shown: the first at $z = 0.1 \pm 0.02$ with $\lambda = 16.2$ and the second at $z=0.56 \pm 0.05$ with $\lambda = 15.8$. 

Fig.~\ref{fig:charts} shows finding charts for these two cluster candidates, which identify potential member galaxies and those with spectroscopic redshifts. The spectroscopic redshifts given in the upper image are slightly (by $0.03$) different from peak redshift. Spectroscopic redshifts of galaxies in the lower image for the low-redshift multiplicity peak are spread from 0.06 to 0.1 because of the way spectroscopic information was included into the cluster search (see Section \ref{sec:specz}). This is an obvious drawback of our method, but it is of importance only at low ($z < 0.12$) redshifts, where other limitations of our method are also relevant (see Section \ref{sec:general}). It is of course preferred to use spectroscopic redshift wherever available as they are more precise.

In the lower image, galaxies are indicated for the two mentioned possible detections: those at $z=0.1$ with light blue and at $z=0.56$ with yellow circles. In this case it is more likely that the X-ray source corresponds to the detection with the higher redshift, as its likely that member galaxies (indicated with yellow circles) show a clear concentration towards the X-ray source, while the galaxies that are contributing to the low-redshift detection obviously do not; this is the case even  though some of these galaxies are very close to the X-ray source. 

Currently the tool does not provide an automated way to select the cluster of galaxies that appears more likely to be associated with the X-ray source in such cases. It has to be identified manually. Experience tells that in many cases the choice is not difficult, taking into account values of $\lambda$, redshift, brightest cluster galaxy (BCG) properties, X-ray extent, and the presence of other extended X-ray sources in the field of view. However, in other cases no decision can be made and X-rays might be emitted by two structures along the line of sight. 

\subsection{Integrated Cluster Finder project code and data}\label{sec:icf_code}
We developed a code implementing the method described above. It is written mainly in Python, and most numerically intense parts are moved to a FORTRAN-90 library. Data are stored in a PostgreSQL database.

\begin{table*}[h]
\centering
\caption{List of catalogues currently used by the  ICF}\label{tab:catalogues}
\begin{tabular}{@{}lcrrr@{}}
\toprule
catalogue & \begin{tabular}[c]{@{}c@{}}Frequency\\bands\end{tabular}& \begin{tabular}[c]{@{}c@{}}Covered \\ Area\\ ($deg.^2$)\end{tabular} & 
\begin{tabular}[c]{@{}c@{}}Number \\ of objects\end{tabular}&\begin{tabular}[c]{@{}c@{}}3XMMe\\clusters\\overlap\end{tabular} \\ \midrule
AllWISE     & W1 (3.4$\mu$m), W2 (4.6$\mu$m) & 41000 & 747,634,026 & 1704 \\
UKIDSS      & JHK      &  4000 &  82,655,526 & 326 \\
SDSS (DR9)  & ugriz   & 14555 & 932,891,133 & 1043 \\ 
CFTHLS-Wide & photo-z &    157 & 35,651,677 & 150 \\ 
CFTHLS-Deep & photo-z &   5.25 &  2,293,851 &  38 \\ 
ALHAMBRA    & photo-z &     4  &    441,303 &  18 \\ \bottomrule
\end{tabular}
\end{table*}

The ICF can in principle use  a variety of data sources: catalogue data stored in the ICF database (which so far includes CFHT deep and wide as well as ALHAMBRA photo-z catalogues) and any VizieR\footnote{\url{http://vizier.u-strasbg.fr/vizier/index.gml}} (or any other Virtual Observatory (VO) compliant catalogue) or output of the new ARCHES cross-matching tool applied to a given catalogue set  \citep{Pineau2015, Pineau2016}. Several Python classes were implemented in the ICF for different types of data sources, and new classes can be easily added. As of now the ICF is highly efficient; its bottleneck is the response time of VizieR server requests.

The set of external catalogues that are currently available for the ICF are listed in Table~\ref{tab:catalogues}. The first column of this table shows the common name of the catalogue. The second column list bands provided by the catalogue or photo-z for photometric redshift catalogues, and the third and fourth columns contain the total sky area (in $deg^2$) and total number of objects in the full catalogue, respectively. The fifth column contains the number of sources from the 3XMMe cluster science-case subset (see section~\ref{sec:3xmme}) that fall into the catalogue area.

For the current work combined data from three large-scale surveys are used in the first place, the SDSS, the UKIDSS (Large Area Survey, LAS), and AllWISE. We use \textit{ugriz} model magnitudes from SDSS, YJHK default magnitudes from UKIDSS (LAS) and W1 and W2 magnitudes from AllWISE (W3 and W4 are too shallow to be used for cluster search). As an alternative  photometric redshifts from CFHT deep and wide catalogues as well as from ALHAMBRA were also used. The photometric data are supplemented by a spectroscopic meta-catalogue that was built from the combination of SDSS (BOSS) and VIPERS \citep{VIPERS} spectroscopic catalogues. Entries in those catalogues are cross-matched with the new ARCHES cross-match tool \citep{Pineau2015, Pineau2016}. The output of the tool is the list of possible matches with match probabilities indicated for each combination. The combination of sources from different catalogues with the highest probability was always used for cluster search. There might be confusing cases with two or more combinations having similar match probabilities, but these are rare and are not expected to affect the cluster search. 

We use the following criteria to select galaxies from the three main catalogues (in parenthesis we list constraints as used in the VizieR web-masks):

\begin{description}
 \item[SDSS] only select primary (mode=1), good (Q=3), galaxies (cl=3) with non-zero and positional errors that are not too large (e\_DEJ2000~\textless~100).

 \item[UKIDSS] select non-duplicate (m=1) sources with at least one frame (nf~\textgreater~0).

 \item[For AllWISE] select not very extended (ex~\textless=~1) sources with high signal-to-noise (snr1~\textgreater~4) and having non-zero errors in position (eeMaj~\textgreater~0) and W1 magnitude (e\_W1mag~\textgreater~0). We add a further positional error of 0.01" in quadrature to take possible systematic errors into account.
\end{description}

\subsection{Calibration of the redshift-colour relation}\label{seq:colourredshift}

Passive red galaxies at a given redshift form a so-called red sequence \citep[see e.g.][]{Koester2007}, that is a (linear) relation between a colour and a magnitude of the galaxies. Red sequences were found up to high redshifts, which is a property that qualifies them for cluster identification and redshift measurement over a wide redshift range. Red sequences at a given redshift typically have a small slope that is ignored here. An obvious effect of this omission is that the spread in a given colour of red galaxies at a given redshift (i.e.~the diagonal elements of the covariance matrix) is slightly increased.

There are two possible ways to estimate average colours and corresponding colour-colour covariance matrices as functions of redshift.

The first way is to use some model for a passively evolving red galaxy 
\citep[for example the model from][]{BC2003}. This gives a robust estimate of the colour(s). However, colour-colour covariance matrices have to be generated by varying age and metallicity of models to try to fit the observed data. This is difficult to achieve in a consistent way for a wide redshift range. A further problem is the poorly known redshift evolution of the models. This approach is not followed here for
the two mentioned limitations.

The alternative way to estimate average colours and corresponding
colour-colour covariance matrices as functions of redshift is to extract a redshift-colour relation from available well-calibrated multi-colour data. To this end, we defined a redshift grid ranging from $z=0.02$ to $z=0.8$ with a step-size of $0.02$ After calibration the grid step is reduced to $0.01$ and intermediate values are filled by linear interpolation between points. For each redshift value $z_{\rm bin}$ in the grid we collected colours of galaxies with known spectroscopic redshift close to $z_{\rm bin}$ from spectroscopic surveys. We used a Gaussian mixture model (hereafter GMM)\footnote{\url{http://scikit-learn.org/stable/modules/mixture.html}} to separate red and blue galaxy populations. Then we made a first approximation to the mean colour values $\cavg$ and the covariance matrix $\covariance$ for red galaxies. The functions $\cavg$ were additionally smoothed above $z=0.5$ by applying spline interpolation.

The next step is to take photometric errors of the calibration data into account. We carried this out by applying the error-corrected Gaussian mixture model \citep[see][]{ecGMM}. This model was only developed for one colour, so it is run  for each colour separately here. Colours were taken from a first approximation and then frozen, allowing for different weights for the blue and red populations and for variable dispersions. We then obtained for a colour, $C_j$, a new dispersion of red galaxies colour $\delta_j^2$. We now have to replace a diagonal element $\mathbf{C}_{int,jj}$ representing the colour dispersion in the covariance matrix with a new value $\delta_j^2$, effectively shrinking the covariance ellipsoid in the $j$th dimension. This is carried out consistently by multiplying the $j$th row and the $j$th column of $\mathbf{C}_{int}$ by $\delta_j/\sqrt{\mathbf{C}_{int,jj}}$.
As a result of this procedure one obtains the mean colour values $\cavg$ and covariance matrices $\covariance$.

\begin{figure}[ht]
  \includegraphics[width=0.47\textwidth, angle=0,clip=]{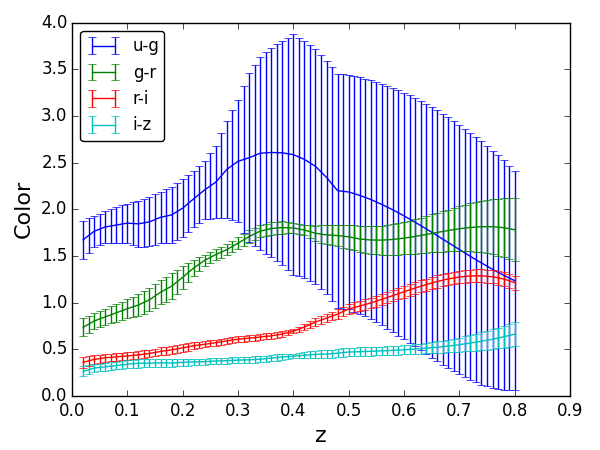}
  \includegraphics[width=0.47\textwidth, angle=0,clip=]{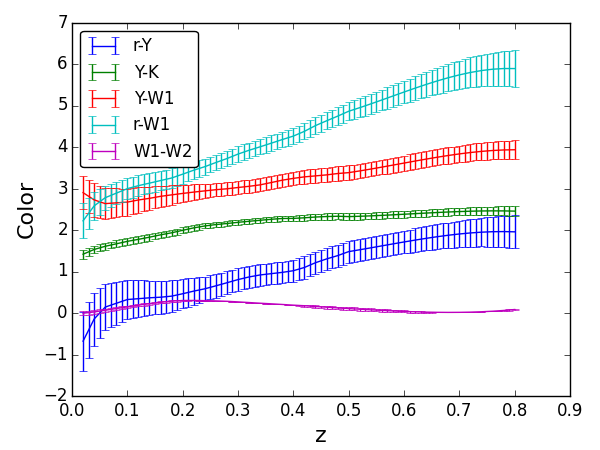}
  \caption{Redshift-colour relation obtained from the calibration for colours from the SDSS catalogue 
  (upper) and the \XMatch combination (lower). Error bars 
  illustrate the spread in this colour and were derived from diagonal 
  elements of the covariance matrix $\covariance$.}\label{fig:z_colour}
\end{figure}

For the SDSS-UKIDSS-AllWISE catalogue combination various colours $\cavg$ are shown in Fig.~\ref{fig:z_colour}. It can be seen that at different redshifts, the various redshift-colour relations have different slopes, thus providing different amounts of information concerning the redshift. For example, $g-r$ changes rapidly below $z < 0.4$ but shows little variation above that redshift. The colour $r-i$ displays the opposite behaviour. Therefore some authors used $g-r$ to estimate redshifts for $z < 0.4$ and $r-i$ for $z > 0.4$ \citep{Ascaso2011, Erfanianfar2013}. The lower panel of Fig.~\ref{fig:z_colour} shows why optical-to-infrared colours are so useful for the ICF. The colours $r-Y$ and $r-W1$ change almost linearly over the complete redshift range considered here. The UKIDSS alone (see $Y-K$ colour) or AllWISE alone (see $W1-W2$ colour) are not very useful as colours in each catalogue does not change much over the redshift range, but their combination provides the colour $Y-W1$ that can be used as it changes monotonically over the redshift range, although the slope is small. For redshifts above $z \approx 0.6$ IR colours become more and more important, as the 4000 \AA~break moves closer to the infrared domain.

In the current implementation it is not necessary for a galaxy to have measured magnitudes for all bands, and thus all colours; a result is generated even with just one colour, with of course lower precision. We cannot however use AllWISE alone on areas not covered by SDSS, for two reasons: first, the $W1-W2$ colour is not a good redshift proxy in the redshift range that is of interest here, second, as shown below in Section \ref{sec:limitations}, $W2$ is not deep enough to give reliable results at $z > 0.3$.

\subsection{Galaxy selection and magnitude cuts}
Within a circle of a radius $\tilde{R}(z)$ (the minimum between $R = 8'$ and $R_s$ -- the distance corresponding to 1 Mpc projected distance at redshift $z$) around a given position, galaxies were selected at each redshift that are in the range $m_*(z) - 3 < m < m_*(z) + 2,$ where 
$m_*(z)$ is the magnitude of a galaxy with a mass of $2\times 10^{11} M_{\odot}$. For this selection we used the SDSS $r'$-band magnitude, if the object had SDSS data and the AllWISE $W1$ magnitude otherwise. The values of $m_*(z)$ were taken from \cite{BC2003}. We call the magnitudes of the two chosen bands the "main" magnitudes, and use these  for the computation of the Schechter function (see Eq.~\ref{eq:luminosity}).

We ran a test to check the sensitivity of our method to the chosen magnitude limits. We took the final ICF sample (see section \ref{sec:icfcatalogue}) and ran a modified ICF with magnitude limit set to $m_*(z) - 3 < m < m_*(z) + 1$. We missed 19 clusters as compared to the ICF sample (i.e. less than 4\%) and detected 9 "new" clusters, mainly at low redshift and with low multiplicity $\lambda$.

\begin{figure}[th]
  \includegraphics[width=0.5\textwidth, angle=0,]{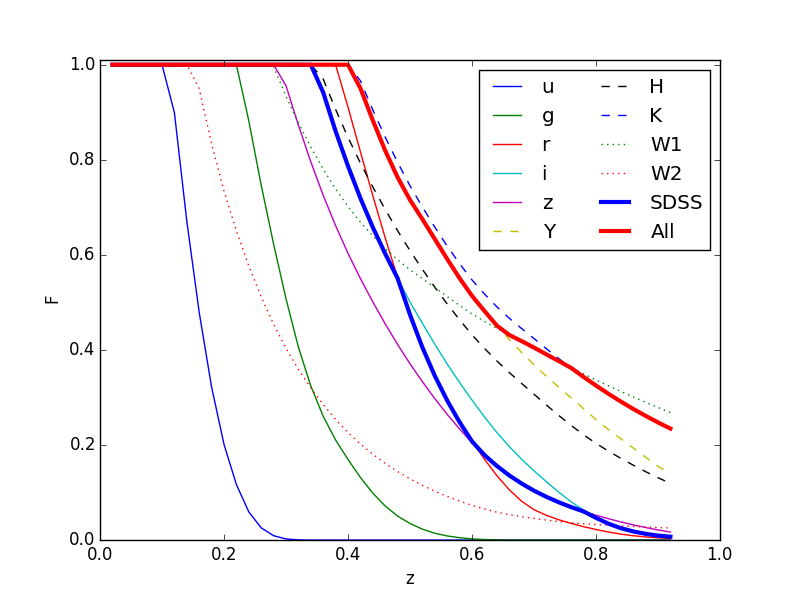}
  \caption{Fraction of all galaxies in the catalogue brighter than the completeness limit within the range $m_*(z) - 3 < m < m_*(z) + 2$ for SDSS (thin solid lines), UKIDSS (dashed lines), and AllWISE (dotted lines) bands. Thick lines show fractions of galaxies with at least 2 bands (i.e. at least 1 colour) for SDSS alone (blue) and all 3 catalogues combined (red).}\label{fig:depth}
\end{figure}

\subsection{Limitations of catalogues}\label{sec:limitations}
Figure~\ref{fig:depth} shows fractions of all galaxies in the range $m_*(z) - 3 < m < m_*(z) + 2$ in the catalogue that are at the same time brighter than the completeness limits for SDSS, UKIDSS, and AllWISE as a function of redshift.
 A magnitude limit was used for UKIDSS instead of a completeness limit \citep{SDSS_main, UKIDSS, AllWISE}, and the 90\% completeness limit should be smaller, thus the dashed curves should move to the left. Thick lines show fractions of galaxies with at least two bands (i.e. at least one colour) for SDSS alone (blue) and all three catalogues combined (red). In the absence of UKIDSS data, in fact, we cannot go much deeper than the SDSS alone, as the AllWISE W2 band is relatively shallow. Nevertheless, there is still an advantage to using AllWISE data, as one gets more colours and thus a better redshift estimate at intermediate redshifts. We decided to place an upper redshift limit for the cluster search for SDSS-UKIDSS-AllWISE at $z=0.8$. Photo-z catalogues used here are much deeper than SDSS, so an upper limit was placed at $z=1.4$.

It was decided not to correct for the incompleteness as the exact fraction of missed galaxies is uncertain for UKIDSS and AllWISE; their depth varies from point to point on the sky.

\subsection{Determination of the background galaxy density}\label{sec:background}

The background function $b(z, x)$ is defined as the distribution of background galaxies in magnitude $m_{main}$ and $p_\nu$ for each value of redshift $z$. As mentioned above, for the SDSS-UKIDSS-AllWISE combination of data two "main" magnitudes are used, and hence two background functions were determined for each value of $z$, magnitude $m_{\rm main}$ and probability $p_\nu$.

For the determination of the background maps 4000, circular fields with 8' radii were analysed at positions that were randomly chosen in the SDSS area. These fields were treated with the cluster finder in the same manner as if they were X-ray sources. The only difference was that rather than using galaxies within the certain radius $\tilde{R}(z)$ (see Sec.~\ref{sec:lambda}), all galaxies within a radius of 8' from the chosen position were used. The number of galaxies in each bin of $m_{main}, p_\nu$ was divided by the total area of the selected background fields, which is $4000 \times \pi (8')^2$.

For photometric redshift catalogues, we built the background function $b(z, x)$ as the distribution of all galaxies in the catalogue in magnitude $m_{main}$ for each value of redshift $z$, divided by the total catalogue area.

\section{Tests of the tool}
\label{sec:tests}
\subsection{Influence of infrared data}

\begin{figure}
  \centering
  \includegraphics[width=0.47 \textwidth,angle=0]{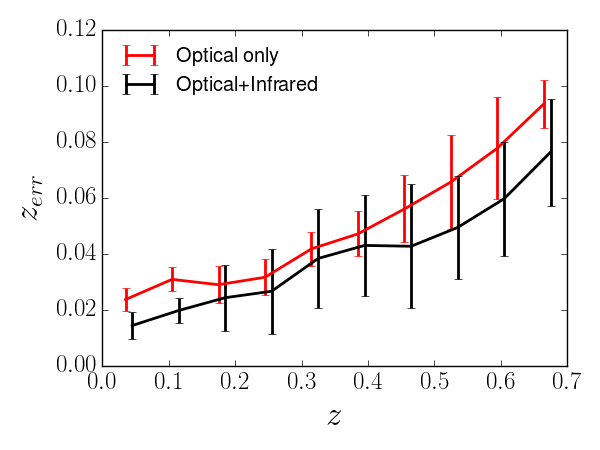}
  \caption{Mean redshift error as a function of redshift for optical data alone (red curve) and combined optical-infrared data (black curve). See text for details.}\label{fig:oandir}
\end{figure}

We explored the impact of the use of the infrared data by comparing redshift accuracies of cluster redshifts that were obtained with and without the infrared catalogues. We used our ICF catalogue for this test (see Section \ref{sec:icfcatalogue}). Detected clusters were grouped into ten bins in the range $0 < z < 0.7$. The result is shown in the Figure \ref{fig:oandir},
in which we plot average redshift error for each redshift bin. 
The red curve corresponds to redshifts obtained with SDSS data alone, and the black curve to redshift obtained with the \XMatch combination.
The error bars indicate a spread in redshift errors within each bin. The small horizontal offset was added to the red line for clarity.
Redshift errors were found to be approximately 20 percent smaller when we included infrared data.

\subsection{Dependence on method parameters}
\label{sec:paramtest}

\begin{figure}
  \centering
  \includegraphics[width=0.47 \textwidth,angle=0]{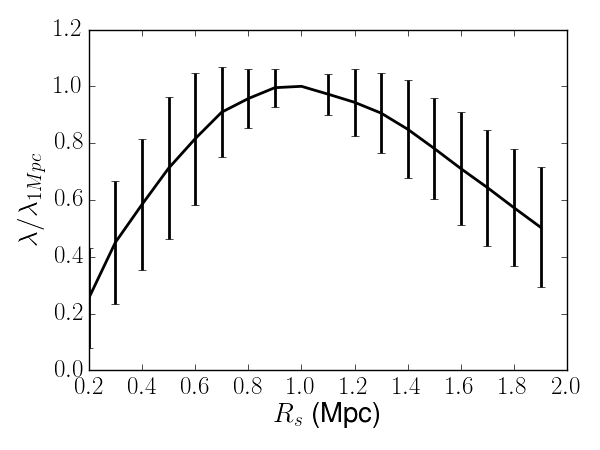}
  \caption{Average $\lambda/\lambda_{\rm 1Mpc}$ as a function of $R_s$ for 50 clusters within $0.30 < z < 0.45 $ and $\lambda_{\rm 1Mpc} > 10$.}\label{fig:param_r}
\end{figure}

\begin{figure}
  \centering
  \includegraphics[width=0.47 \textwidth,angle=0]{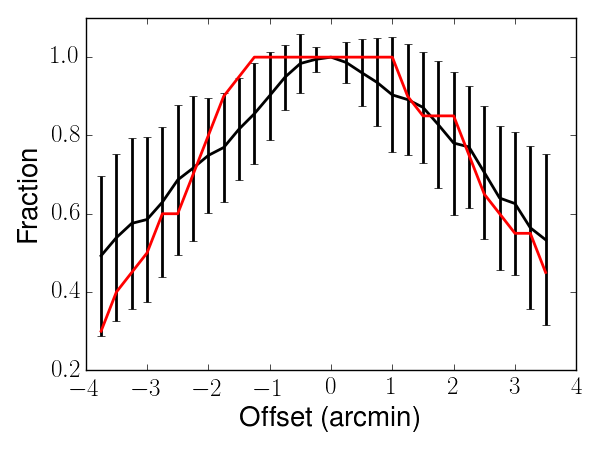}
  \caption{Average $\lambda/\lambda_0$ as a function of offset (black line) and fraction of clusters still detected with a given offset (red line).}\label{fig:param_offset}
\end{figure}

We tested how stable our method is in two ways. First we started changing the extraction radius $R_s$ (see Section \ref{sec:lambda}) from 0.2 to 2 Mpc (the value used by the ICF is fixed to 1 Mpc). We took 50 clusters from the reference catalogue (Takey sample of clusters) for which two conditions are satisfied: $0.3 < z < 0.45$ and $\lambda > 10$. We took this redshift range so that all our clusters were at comparable distances, so the linear scale would not change much. The result of the test is shown in the Figure \ref{fig:param_r}, where we show average values of $\lambda$ normalized to the default value with $R_s = 1$ Mpc. It can be seen that changing $R_s$ between 0.7 and 1.3 only has a little impact on $\lambda$. Outside this range the measured $\lambda$ is systematically lower. If $R_s$ is much smaller than the actual cluster size, then real cluster members far from the cluster centre are not detected. If $R_s$ is larger than the actual cluster size then we include too much background, thus decreasing $\lambda$.

Our second test  checked how sensible the ICF is to the chosen input position of the cluster. We started changing the assumed cluster positions for the above sample within $\pm 3.5'$. The result of the test is shown in the Figure \ref{fig:param_offset}. It can be seen that the method is stable within $\pm 1.5'$ offsets, which corresponds to 300-500 kpc within the selected redshift range. If the offset is larger than $1.5'$ then the fraction of still detected clusters and their $\lambda$ values decrease rapidly. Therefore we can conclude that our method works best if the offset between the optical cluster and input position is less than $1.5'$.

\subsection{Comparison with reference catalogues}

The performance of the ICF was tested using cluster samples that are available in the literature.
We were mainly interested in the recovery rate of clusters drawn from various reference catalogues, i.e. the fraction of clusters for which the redshift measured by the ICF is consistent with the redshift provided in the catalogue. 
The reference catalogues used for this exercise are listed in Table~\ref{tab:reference}. They represent a mix of optical and X-ray selected cluster catalogues with redshifts measured either 
from SDSS data (photometric redshifts in most cases) or from dedicated optical follow-up (ROSAT-ESO Flux-Limited X-Ray, REFLEX). The reference catalogues were chosen with the expectation that the clusters to be identified as counterparts of 3XMMe extended sources are somehow similar. Therefore some famous catalogues like the Zwicky and Abell catalogues were not used for this recovery exercise. The clusters listed in those catalogues are typically much more extended than those discovered as single and completely covered entities in single XMM-Newton observations. 

The ICF was run on positions of clusters listed in the reference catalogues. Initially only \XMatch\ data were used as input for this run of the ICF. The measured redshifts $z_{\rm ICF}$ were compared with reference values $z_{\rm ref}$. A cluster was regarded as recovered if the following condition holds: 
$| z_{\rm ICF} - z_{\rm ref} | < \delta z_{\rm ICF} + 0.01$. We added $0.01$ (grid step size) to the redshift error to soften the effect of very small ICF redshift uncertainties in some cases.  The added constant is much smaller than a typical redshift uncertainty in the reference catalogue. Below we briefly comment the results.

\begin{figure}
  \centering
  \includegraphics[width=0.47 \textwidth, angle=0]{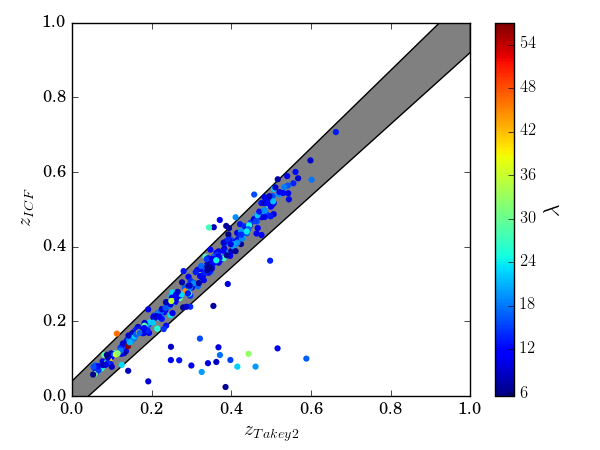}
  \caption{Redshifts estimated by \cite{Takey2} and in this work ($z_{ICF}$). 
Colours reflect the value of $\lambda$. Shaded area indicates 4\% redshift error}\label{fig:takey2}
\end{figure}

\begin{table*}
\centering
\caption{List of test or reference cluster catalogues\label{tab:reference}}
\begin{tabular}{lccccl}
\hline
Cluster catalogue & Number of objects &  \multicolumn{3}{c}{Subset used for testing} & Reference \\
&& z range & Used objects & Recovered & \\
\hline
RedMapper   & 25325 & 0.53-0.55 & 548     & 536 (98\%) & 1  \\
Wen and Han  & 1757  & 0.16-0.8  & 524     & 313 (60\%) & 2     \\
MaxBCG      & 13823 & 0.1-0.3   & 995     & 991 (99\%) & 3 \\
Takey et al.      & 530   & 0.03-0.7  & 515     & 491 (95\%) & 4   \\
REFLEX      & 296   & 0-0.46    & 296     & 284 (96\%) & 5   \\
\hline
\end{tabular}
\tablebib{
(1)~\citet{Rykoff2013}; (2) \citet{Wen2011}; (3) \citet{Koester2007}; (4) \citet{Takey2}; (5) \citet{REFLEX}.
}
\end{table*}

For the sake of simplicity, only the first 1000 clusters were used from the MaxBCG catalogue since this was sufficient for a fair comparison of the success rates between the two catalogues. 
 Only clusters in the highest redshift bin provided (0.53-0.55) were used from the redMaPPer catalogue. The \cite{Wen2011}~sample contains clusters selected in the deep fields so it is much deeper than our upper redshift limit ($z=0.8$), so the test was restricted to clusters in CFHT-wide area with $z < 0.8$. We used the \XMatch data for testing, so one should not expect a very good recovery rate even with this restriction.

\cite{Takey2} compiled a sample of 530 clusters from the common 2XMMi-DR3 and SDSS-DR7 survey area. All those clusters have redshifts ($z_{\rm ref}$) determined from photometric or spectroscopic redshifts of likely member galaxies. All clusters were visually screened so that the resulting catalogue was assumed to be a reliable resource. A subset of those clusters had previously determined redshifts from other surveys ($z_{\rm pub}$). A cluster was regarded as recovered if the above condition was fulfilled for either $z_{\rm pub}$ or $z_{\rm ref}$. Most galaxies in 15 fields of size 8 arcmin around the Takey et al. clusters did not satisfy the selection criteria given in the section \ref{sec:icf_code} and these clusters were thus removed from our sample. The result is shown in Figure~\ref{fig:takey2} in $z_{\rm ref}$ versus $z_{\rm ICF}$. Out of 515 input clusters, 491 (95\%) were recovered with the above criteria.
Discrepancies are caused by the fact that clusters in this catalogue are relatively faint and their redshifts are less certain owing to less photometric information. A concentration of outliers at the lower part of the plot is caused by low-redshift detections made by the ICF. These detections are caused by the presence of a low-redshift cluster in the XMM field that was used by \cite{Takey2} to search for clusters; such clusters are often the primary observation target. 
Another example of such a problem is shown in the lower part of image \ref{fig:charts} and discussed in detail in Section \ref{sec:peakfinding}.
This problem is relevant for the ICF cluster search based on 3XMM (see Section \ref{sec:icfcatalogue}).

\begin{figure}
  \centering
  \includegraphics[width=0.47 \textwidth, angle=0]{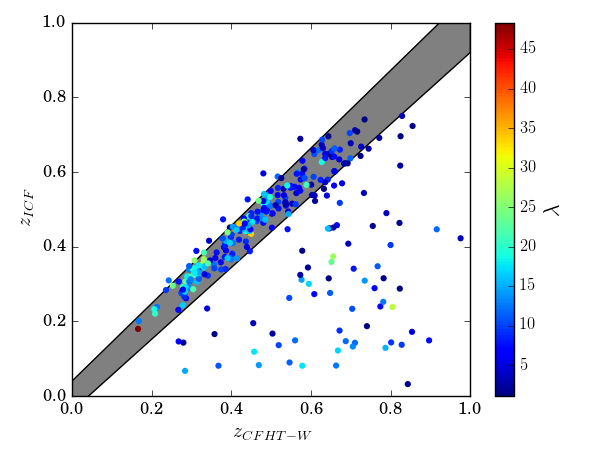}
  \includegraphics[width=0.47 \textwidth, angle=0]{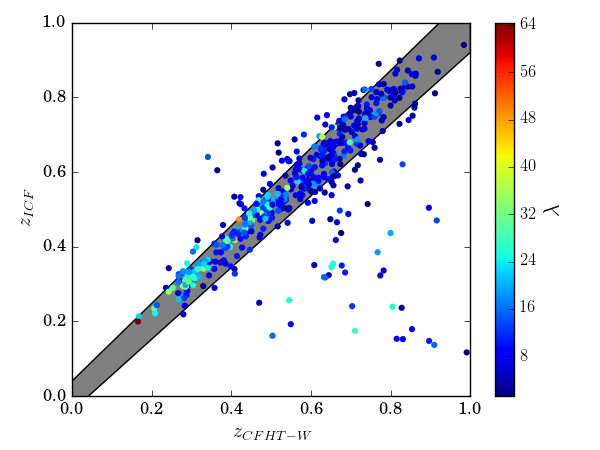}
  \caption{Redshifts estimated by \cite{Wen2011} and in this work ($z_{ICF}$). Upper plot was produced with \XMatch data, whereas the lower one was produced with CFHTLS photometric redshifts data. Colours reflect the value of $\lambda$. See text for more details.}\label{fig:cfht-w}
\end{figure}

As it is shown in Table~\ref{tab:reference}, 95\% to 99\% of clusters were recovered by the ICF for four out of five reference samples. A lower recovery rate was obtained for Wen and Han sample owing to a much higher average redshift in their catalogue. This cluster catalogue was constructed with CFHTLS data, which are much deeper than SDSS. If we consider only a subsample with $z < 0.5$ than 88\% of clusters were recovered. As a comparison, we tested the same sample using photometric redshifts from CFHT-W catalogue as input. Results are shown in Fig.~\ref{fig:cfht-w}. It is obvious that we improve the recovery rate substantially (up to 84\% recovered clusters) by using higher quality data .

The lower recovery rate (96\%) for the REFLEX dataset is partially caused by discrepant redshifts in the REFLEX catalogue (this was revealed by comparing REFLEX to MaxBCG and redMapper) or by low redshifts of the reference catalogue. As described above, the ICF performs poorly in detecting clusters at $z < 0.12$. If the input is restricted to clusters with redshift $z > 0.12$ 161 out of 165 REFLEX clusters (over 97\%) are recovered.

\section{Completeness and false positives fraction}\label{sec:fom}
An important step in cluster finding is the estimation of completeness (i.e.~what is the fraction of clusters that are not detected) and of the fraction of false positives (i.e.~what is the chance of detecting a cluster that does not exist or is not related to an X-ray source).

We define $F(r)$ as the cumulative function of the NFW profile (Eq.~\ref{eq:nfw}). Then we define the inverse cumulative NFW function $F^{-1}(t): F^{-1}(F(r)) = r$. Now we define $\rnfw = F^{-1} \left( \frac{\sum_{i=1}^{n} F(r_i)}{n} \right)$. This value gives a characteristic cluster size. For a purely random set of points, we expect $\rnfw = 0.5$ on the average. 

The same 4000 random fields that were used for the background determination (see Sec.~\ref{sec:background}) were used to determine the false positives rate. The ICF was run for these fields (just as if there was an X-ray source). The main goal of this experiment was to estimate the probability of a cluster ''detection'' in a case when the X-ray source is spurious or not related to a cluster (e.g.~two unresolved point sources or a nearby galaxy or similar). These detections were compared to the detections of clusters from the reference sample.

\begin{figure}
\myimage{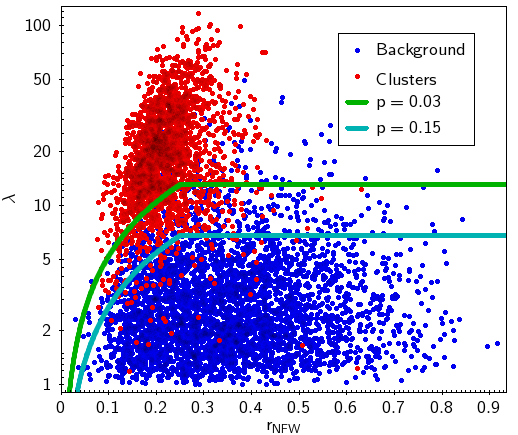}
\caption{$\rnfw$ vs $\lambda$ for MaxBCG, redMapper, Takey and REFLEX clusters (red dots) and background fields (blue dots). Lines indicate constant $p_{sp}(FOM)$ with values of 3 and 15 percent.} \label{fig:rlambda}
\end{figure}

The detections of the two datasets presented above (reference clusters and background fields) are shown in the $r_{NFW}$ - $\lambda$ plane in Fig.~\ref{fig:rlambda}. Recovered clusters from the reference catalogues (see Sec.~\ref{sec:tests}) are shown with red dots and detections from background fields are indicated with blue dots. The way we choose cluster members (see Section \ref{sec:peakfinding}) selects galaxies closer to the centre preferentially, therefore the  average $\rnfw$ is somewhat smaller than $0.5$ even for
a spurious detection, as can be seen in the Figure \ref{fig:rlambda}.

The cores of the two datasets are well separated in the plane chosen but they also have some overlap. 
It may well be that some positions from the randomly selected set might indeed contain a real cluster so that the expected distribution of a pure background non-cluster sample would be confined to smaller values of $\lambda$. On the other hand, some of the sources from the reference samples might be spurious and thus be located at rather low values of $\lambda$.

Here we want find a separation curve in a way that as many background detections as possible
are kept below this curve and as many ``real'' clusters, i.e.~clusters drawn from the reference samples, as possible are above it. Luckily the exact shape of the curve has little impact on the result;  after testing exponential, power-law, and linear separation lines we chose the broken-line shape as the most simple, but still reliable, solution.

We want to produce a probability, by means of these separation lines, for a given detection to be spurious as a function of $\lambda$ and $\rnfw$. This is carried out by introducing a figure of merit (FOM) as: 
\begin{equation}
  {\rm FOM}(\lambda, \rnfw) = \left\{
   \begin{array}{ll}
     \frac{\lambda}{\rnfw},& \rnfw < 0.25 \\
     4 \lambda, & \rnfw \ge 0.25.
   \end{array}\right.
        \label{eq:fom}
\end{equation} 
The case for $\rnfw \ge 0.25$ is motivated by the fact that there seems to be little dependence of background detections distribution in $\lambda$ on $\rnfw$, so we can use ${\rm FOM}(\lambda, \rnfw) = {\rm FOM}(\lambda, 0.25)$.

We then define $p_{sp}({\rm FOM})$ as a fraction of detections from the background fields with ${\rm FOM}' > {\rm FOM}$, i.e. above the line defined by the equation~\ref{eq:fom}. It can be approximated by the following equation:
\begin{equation}
  p_{sp}{\rm (FOM)} = 4892 ({\rm FOM}(\lambda, \rnfw)^2 + 287.178)^{-1.5},
\end{equation}
where FOM is calculated via Eq.~\ref{eq:fom}. It is not very precise at higher values, and can even produce $p_{sp}(\rm{FOM}) > 1$, but clusters with high $p_{sp}{\rm (FOM)}$ are rejected anyway as likely to be spurious.

The completeness $f_{com}(x)$ is defined as the fraction of clusters for which ${\rm FOM} > x$ and the false positives rate (or the spurious detection fraction) $f_{sp}(x)$ as the fraction of background detections for which ${\rm FOM} > x$.

We stress that completeness was derived from a set of catalogues of clusters that can be biased and can contain spurious detections as well.
On the other hand,  there might be some real clusters, for example a low-redshift, large Abell cluster or a portion thereof, even among background detections. So there still might be some detections in the final catalogue that are not in fact related to the X-ray source used. We adopted a false positives fraction of 15 percent. This does not mean that the contamination of the produced catalogue by cluster detections that are unrelated to the X-ray source are as high because we use the additional information that there is an extended X-ray source at this location.

\section{Integrated cluster catalogue}\label{sec:icfcatalogue}

\subsection{Cluster catalogue production}
\subsubsection{Data selection}\label{sec:3xmme}
The input catalogue of cluster candidates used here is based on 3XMM-DR5 \citep{3XMMDR5}. Within ARCHES this catalogue was filtered and enriched with flags for possible scientific applications and the column density of cold interstellar matter in the direction of the source. This catalogue version is called 3XMMe, where the 'e' stands for 'enhanced'.\footnote{You can find a detailed description of 3XMMe catalogue at \url{http://www.arches-fp7.eu/images/PDF/3XMMe_catalogue_v2.1.pdf}}

 X-ray sources other than clusters may appear as X-ray extended and they clearly do not fall under the category to be studied in this paper, namely nearby galaxies, solar system objects, Supernova remnants, and non-resolved blends of point sources. Some extra measures were taken to discriminate between real clusters and the other sources. Solar system objects did not enter 3XMM, supernova remnants were excluded through the selection of high-galactic latitude fields only, and nearby galaxies were discarded through a careful selection of fields case by case \citep{3XMMDR5}. There is no safeguard against imperfections of the source detection process itself, which is limited by the X-ray optics and the available photon numbers. Quantifying its abilities and limitations would require intensive simulations that are beyond the scope of this paper. 
Entries in 3XMM were thus taken at face value. 

Several cuts and filters were applied to select cluster candidates:
\begin{enumerate}
\item Observations with high background, hotspots, and corrupted mosaic mode data were removed.
\item Low exposure ($< 5$ks) observations were removed.
\item $0 < EP\_EXTENT < 80$ arcseconds. This only considers detections with real extent, that is below the upper limit of 80 arcsecs imposed in the source detection step within the standard  XMM-Newton pipeline processing.
\item $EP\_EXTENT\_ERR < 10$. This excludes poorly constrained extent values.
\item The  galactic  latitude must  satisfy  the  constraint  $|b_{II}|  >  20.3$ degrees.
\item $EP\_9\_DET\_ML >10$. This demands  a  minimum  detection  likelihood value of 10 in band 9 (XID band = 0.5-4.5 keV).
\item $SUM\_FLAG < 2$. This excludes  manually  flagged  detections  and  also detections with $sum\_flag=2$ ; these are generally detections that are extended and close to other sources or within the envelopes of other extended sources.
\item $4 < offaxis < 12$ arcmins. The off-axis angle for the detection is measured, in arcmins, from the spacecraft bore sight for the observation. The lower boundary is intended to mitigate against the inclusion of extended target objects, while the upper cuts away low-completeness regions at the edges of the field of view.
\end{enumerate}

The various selection steps resulted in a preliminary input catalogue of 1704 extended X-ray sources in 3XMMe, which are all regarded as potential clusters of galaxies. On top of the constraints that were applied to construct 3XMMe further constraints were applied for cluster finding with the ICF:
\begin{enumerate}
\item Select only sources with SDSS coverage; this leaves 1043 X-ray sources in the catalogue.
\item Remove sources that fall into a field that have bright stars or very bright galaxies (like M31 or Magellanic clouds) in the field of view or the very vicinity; a further 186 sources were thus removed, leaving 857 X-ray positions in the input catalogue.
\end{enumerate}

The SDSS, UKIDSS, AllWISE, and spectral data were retrieved for the fields associated with the remaining X-ray sources. The data were then cross-matched with the new XMatch tool \citep{Pineau2015, Pineau2016}. Where possible, we also collected the photometric redshift data from CFHT deep and wide catalogues and ALHAMBRA; this was possible for 135 unique sources, and   SDSS data was available for some of these as well.

We removed from the analysis all fields that had less than two-thirds of their area covered by survey; otherwise our data would be highly incomplete and cluster detection would be unreliable. This removed fields with bad photometry or those close to the survey borders. We were thus left with 820 sources with SDSS data. As for photometric redshifts data, 124 fields had proper coverage, but only 10 of these fields were not contained in SDSS.

The ICF was run on all these data. Cluster candidates were considered as detected if $p_{sp}(\lambda, \rnfw) < 15\%$.
For each detection the following information is stored:
\begin{itemize}
\item X-ray source coordinates
\item Estimated redshift and its error
\item Multiplicity value $\lambda$
\item Weighted radius $\rnfw$
\item Probability for the detection to be spurious $p_{sp}$
\item Distance of the BCG from the cluster centre
\item Difference between the magnitude of the BCG and $m_*(z)$\end{itemize}

We also saved information for all likely members of each cluster in a separate table (id, position, magnitudes in all bands and, if available, spectroscopic redshift).

\subsubsection{Duplicates and matches to known clusters}

A few further steps need to be performed in constructing the 3XMMe-ICF cluster catalogue.

(a) The ICF was applied using both options (photoz and colour-redshift relation) to search for galaxy overdensities around the given input positions. This could have led to more than one detection of the same source if the position was~within the CFHT and the SDSS area, for example. Only one of the detections was chosen and documented for the construction
of confirmed clusters from 3XMMe. The detection with the higher value of $\lambda$ was always chosen, which typically was the detection from CFHTLS. This contains photometric redshifts for both red and blue galaxies and the multiplicity $\lambda$ even at the same limiting depth of the input data is expected to be systematically larger. Therefore if cluster is detected in \XMatch and photometric redshift data, then the latter detection is chosen in most cases. 

(b) There are few cases when two or three sources from the input X-ray source list appear close to each other on the sky. We get detections with similar redshifts and multiplicities for most of such pairs or groups. This might be due to one of three possibilities: 1) one of the X-ray sources is in fact not related to the cluster, 2) the clusters detected are in the merger state and have a complex X-ray structure, or c) there are two clusters with different redshifts that are nearly aligned along the line of sight. Such cases are marked in the catalogue and have to be treated with care. In such cases, we select a detection we think is most likely to be related to the X-ray source, and flag all other detections as possible duplicates.

(c) In order to give reference to possible previous detections and identifications of the same source, a meta-catalogue of clusters is created. It contains more than 50000 cluster redshifts from the catalogues listed in Table~\ref{tab:reference}. 
This catalogue was supplemented by the Abell and Zwicky catalogue \citep[see][]{Abell} containing 10411 objects. Many clusters are listed there several times, as they were detected in different catalogues. Matching is carried out using the criteria $d < R(z) + R_{\rm ref}$, where $d$ is the angular distance between the detected and reference clusters, $R(z)$ is the angular size corresponding to 1 Mpc comoving size at cluster redshift $z$, and $R_{\rm ref}$ is a reference cluster radius (taken from the reference catalogue data whenever possible, and otherwise fixed to 1 Mpc at the cluster redshift). It is important to note that matching entries between the 3XMMe-ICF catalogue and the reference meta-catalogue does not necessarily mean that these are the same clusters. The example discussed in Fig.~3 is typical for such a case; a new XMM-Newton discovered cluster may be located behind an Abell cluster. Hence a simple positional match between catalogue entries is not sufficient but the redshifts have to be compared as well and need to be matching. It is not easy to define a standardized procedure since not all reference catalogues give, for example a redshift error. In any case, we do not change the redshift found by the ICF but document the existence of potentially matching clusters with their redshift from other cluster catalogues.

(d) Finally the NASA Extragalactic database (NED) database\footnote{\url{http://ned.ipac.caltech.edu/}} is queried to search for known clusters. The search radius was fixed to 8 arcmin. 
The information that was retrieved from the NED for all found matching clusters is included in the catalogue.
\begin{figure}[ht]
\myimage{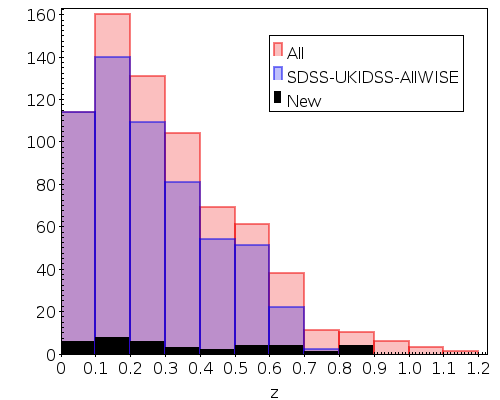}
\caption{Redshift distribution for detected clusters. The red histogram shows non-duplicate clusters detected with all methods; the  blue histogram shows only those detected with SDSS-UKIDSS-AllWISE combination; and the black histogram shows new clusters.}\label{fig:z_distribution}
\end{figure}

\begin{figure*}[!ht]
      \myimagelarge{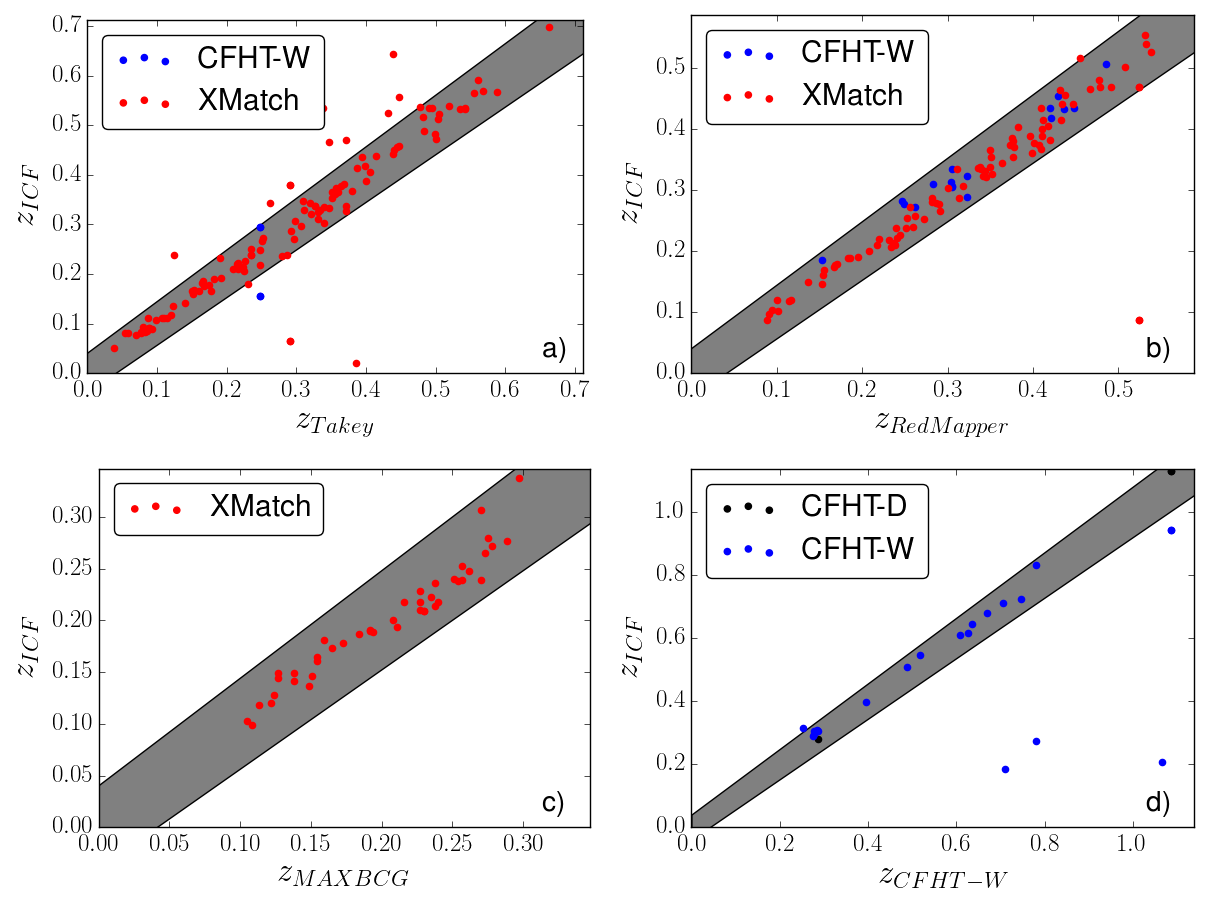}
\caption{Redshifts for detected clusters from other catalogues: a) \citet{Takey2}, b) \citet{Rykoff2013}, c) \citet{Koester2007}, and d) \citet{Wen2011}. Colour of points indicate data used by ICF to detect clusters. Shaded region corresponds to $4\%$ error in redshift.}\label{fig:cross_match}
\end{figure*}

\subsection{Integrated Cluster Finder run results}

The cluster catalogue contains 708 detections on 488 unique X-ray sources. Of those 348 have at least one member with a spectroscopic redshift measured. Only 38 detections do not have a counterpart in NED or the reference catalogue. The redshift distribution of reliable and unique cluster detections is shown in figure~\ref{fig:z_distribution}. Using catalogues with photometric redshifts  we get more detections not only at very high but also at intermediate redshifts. This is caused by the fact that $\lambda$ values for detections based on photometric redshift catalogues are systematically larger and that photometric redshift catalogues cover some parts of the sky observed by XMM-Newton but not covered by SDSS. When the ALHAMBRA catalogues were used in cluster detection they do not reveal any new detection that was not also discovered with the other catalogues.

\begin{figure} 
\centering
\myimage{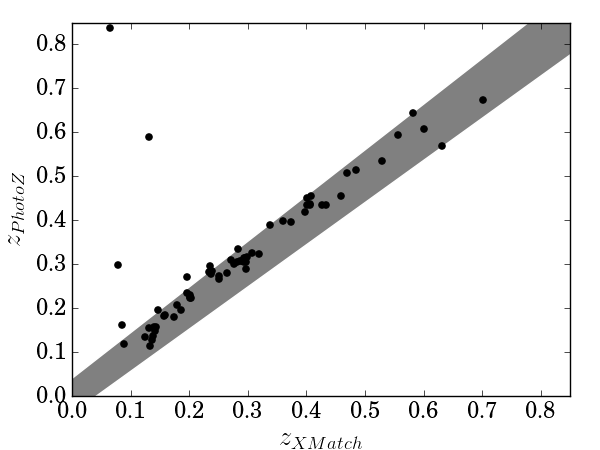}
\caption{Redshift determination with \XMatch and with CFHT or ALHAMBRA photo-z data.}
\label{fig:compare_z}
\end{figure}

Figure~\ref{fig:cross_match} shows redshift comparisons for ICF detections that enter the catalogues listed in Table~\ref{tab:reference}. As before, it can be seen that ICF sometimes measures a much lower redshift than the redshift provided by a reference catalogue. This is caused in most cases by an overlap with a large nearby cluster, which was the main target of the XMM-Newton observation. In the preparation of this diagram, the following method was used for matching clusters from each of the two catalogues in consideration for each panel. The separation between the cluster position used by the ICF, which is equivalent to the X-ray source position, and cluster position from the reference catalogue was requested to be less than 1 Mpc in comoving distances for redshifts reported by both ICF and the reference catalogue. This ensures that a low-redshift ICF detection is not matched with an offset high-redshift reference cluster and vice versa.

Another test is to compare detections of the same cluster made with \XMatch and with photo-z data. The result for 67 such detections is shown in the Figure~\ref{fig:compare_z}. Apart from a few disagreements at low redshifts the correlation is very good. 
Low-redshift outliers are likely caused by the fact that a cluster detected with (more sensitive) photo-z data was simply not detected in \XMatch and, at the same time, another detection with low redshift appeared.

\section{Discussion}\label{sec:discussion}

We have developed a tool to search for galaxy clusters across multiple catalogues involving optical and  infrared data as well as photometric and spectroscopic redshifts. The main purpose for the development of the tool was the search for, and the confirmation and the redshift determination of, clusters drawn as extended sources from the 3XMMe X-ray source catalogue. The reliability of the tool was shown by testing its performance against established cluster catalogues in the literature. We also provide means to filter out spurious detections.

Upon acceptance of this paper an open service will be made available to the community that will facilitate running the ICF on arbitrary user-defined positions. It is an on-line tool hosted by the Observatoire Astronomique de Strasbourg to be accessed using the  URL \url{http://serendib.unistra.fr/icf}.

The cluster catalogue, which is generated on the basis of the final version of the enhanced version of 3XMM called 3XMMe, serves as the starting point for its scientific exploitation. The catalogue is published via the ARCHES database\footnote{\url{http://xcatdb.unistra.fr/3xmmdr5/archesindex.html?mode=catalog}}. It gives information concerning all detected clusters together with a diagram showing the multiplicity $\lambda(z)$ plots (similar to Figure~\ref{fig:lambda_example}). Combined SDSS and X-ray finding charts similar to Figure~\ref{fig:charts} are provided along with lists of member galaxies. An investigation of this new resource was started to further study the $L_X - T_X$ cluster scaling relation between between the temperature and luminosity of the hot intracluster medium. 

We foresee and plan the following extensions and future applications of the tool. 
Right now full functionality is given in the northern hemisphere only because of the lack of suitable wide-field surveys in the south while developing the ICF. This situation will change in the near future with VO-compliant releases of data from, for example~the Dark Energy Survey (DES) \citep{DES}, the Panoramic Survey Telescope and Rapid Response System (Pan-STARRS) \citep{PanSTARRS}, ATLAS \citep{ATLAS}, and the VISTA Hemisphere Survey (VHS) \citep{VISTA}. Correspondingly new colour-redshift relations need to be calibrated for those data and a new download module needs to be written. 
The ICF is a tool that  on input only needs a position on the sky. 
In the context of this paper, input positions were those of the extended sources in 3XMMe\footnote{\url{http://www.arches-fp7.eu/images/PDF/3XMMe_catalogue_v2.1.pdf}} \citep{3XMMDR5}, but there is no restriction that would prevent using it with other source input, for example future XMM-Newton data releases or source catalogues of the upcoming extended Roentgen Survey with an Imaging Telescope Array (eROSITA) mission  with a huge need for optical follow-up~\citep{merloni+12}.

\begin{acknowledgements}
This work has made use of data/facilities from the ARCHES project (7th Framework of the European Union nº 313146). \newline

Francisco~Carrera acknowledges partial financial support by the Spanish Ministry of Economy and Competitiveness through grants AYA2012- 31447 and AYA2015-64346-C2-1 \newline

Funding for SDSS-III has been provided by the Alfred P. Sloan Foundation, the Participating Institutions, the National Science Foundation, and the U.S. Department of Energy Office of Science. The SDSS-III web site is http://www.sdss3.org/.

SDSS-III is managed by the Astrophysical Research Consortium for the Participating Institutions of the SDSS-III Collaboration including the University of Arizona, the Brazilian Participation Group, Brookhaven National Laboratory, Carnegie Mellon University, University of Florida, the French Participation Group, the German Participation Group, Harvard University, the Instituto de Astrofisica de Canarias, the Michigan State/Notre Dame/JINA Participation Group, Johns Hopkins University, Lawrence Berkeley National Laboratory, Max Planck Institute for Astrophysics, Max Planck Institute for Extraterrestrial Physics, New Mexico State University, New York University, Ohio State University, Pennsylvania State University, University of Portsmouth, Princeton University, the Spanish Participation Group, University of Tokyo, University of Utah, Vanderbilt University, University of Virginia, University of Washington, and Yale University.

This work is based in part on data obtained as part of the UKIRT Infrared Deep Sky Survey.

This publication makes use of data products from the Wide-field Infrared Survey Explorer, which is a joint project of the University of California, Los Angeles, and the Jet Propulsion Laboratory/California Institute of Technology, and NEOWISE, which is a project of the Jet Propulsion Laboratory/California Institute of Technology. WISE and NEOWISE are funded by the National Aeronautics and Space Administration.

Based on observations obtained with MegaPrime/MegaCam, a joint project of CFHT and CEA/IRFU, at the Canada-France-Hawaii Telescope (CFHT), which is operated by the National Research Council (NRC) of Canada, the Institut National des Science de l'Univers of the Centre National de la Recherche Scientifique (CNRS) of France, and the University of Hawaii. This work is based in part on data products produced at Terapix available at the Canadian Astronomy Data Centre as part of the Canada-France-Hawaii Telescope Legacy Survey, a collaborative project of NRC and CNRS.

Based on data from ALHAMBRA Data Access Service the at CAB (INTA-CSIC).

\end{acknowledgements}

\bibliographystyle{aa.bst}
\bibliography{collection.bib}

\begin{thebibliography}{32}
\expandafter\ifx\csname natexlab\endcsname\relax\def\natexlab#1{#1}\fi

\bibitem[{{Abdalla} {et~al.}(2011){Abdalla}, {Banerji}, {Lahav}, \&
  {Rashkov}}]{Abdalla2011}
{Abdalla}, F.~B., {Banerji}, M., {Lahav}, O., \& {Rashkov}, V. 2011, MNRAS,
  417, 1891

\bibitem[{{Ascaso} {et~al.}(2012){Ascaso}, {Wittman}, \&
  {Ben{\'{\i}}tez}}]{Ascaso2011}
{Ascaso}, B., {Wittman}, D., \& {Ben{\'{\i}}tez}, N. 2012, MNRAS, 420, 1167

\bibitem[{{Bleem} {et~al.}(2015){Bleem}, {Stalder}, {de Haan}, {Aird}, {Allen},
  {Applegate}, {Ashby}, {Bautz}, {Bayliss}, {Benson}, {Bocquet}, {Brodwin},
  {Carlstrom}, {Chang}, {Chiu}, {Cho}, {Clocchiatti}, {Crawford}, {Crites},
  {Desai}, {Dietrich}, {Dobbs}, {Foley}, {Forman}, {George}, {Gladders},
  {Gonzalez}, {Halverson}, {Hennig}, {Hoekstra}, {Holder}, {Holzapfel},
  {Hrubes}, {Jones}, {Keisler}, {Knox}, {Lee}, {Leitch}, {Liu}, {Lueker},
  {Luong-Van}, {Mantz}, {Marrone}, {McDonald}, {McMahon}, {Meyer}, {Mocanu},
  {Mohr}, {Murray}, {Padin}, {Pryke}, {Reichardt}, {Rest}, {Ruel}, {Ruhl},
  {Saliwanchik}, {Saro}, {Sayre}, {Schaffer}, {Schrabback}, {Shirokoff},
  {Song}, {Spieler}, {Stanford}, {Staniszewski}, {Stark}, {Story}, {Stubbs},
  {Vanderlinde}, {Vieira}, {Vikhlinin}, {Williamson}, {Zahn}, \&
  {Zenteno}}]{SPTSZ}
{Bleem}, L.~E., {Stalder}, B., {de Haan}, T., {et~al.} 2015, \apjs, 216, 27

\bibitem[{{B{\"o}hringer} {et~al.}(2004){B{\"o}hringer}, {Schuecker}, {Guzzo},
  {Collins}, {Voges}, {Cruddace}, {Ortiz-Gil}, {Chincarini}, {De Grandi},
  {Edge}, {MacGillivray}, {Neumann}, {Schindler}, \& {Shaver}}]{REFLEX}
{B{\"o}hringer}, H., {Schuecker}, P., {Guzzo}, L., {et~al.} 2004, Astronomy \&
  Astrophysics, 425, 367

\bibitem[{{Bruzual} \& {Charlot}(2003)}]{BC2003}
{Bruzual}, G. \& {Charlot}, S. 2003, MNRAS, 344, 1000

\bibitem[{{Clerc} {et~al.}(2014){Clerc}, {Adami}, {Lieu}, {Maughan}, {Pacaud},
  {Pierre}, {Sadibekova}, {Smith}, {Valageas}, {Altieri}, {Benoist},
  {Maurogordato}, \& {Willis}}]{clerc+14}
{Clerc}, N., {Adami}, C., {Lieu}, M., {et~al.} 2014, \mnras, 444, 2723

\bibitem[{{Cross} {et~al.}(2012){Cross}, {Collins}, {Mann}, {Read}, {Sutorius},
  {Blake}, {Holliman}, {Hambly}, {Emerson}, {Lawrence}, \& {Noddle}}]{VISTA}
{Cross}, N.~J.~G., {Collins}, R.~S., {Mann}, R.~G., {et~al.} 2012, \aap, 548,
  A119

\bibitem[{{Cutri} {et~al.}(2014){Cutri}, {Wright}, \& et~al.}]{AllWISE}
{Cutri}, R.~M., {Wright}, E.~L., \& et~al. 2014, Explanatory Supplement to the
  AllWISE Data Release Products,
  \url{http://wise2.ipac.caltech.edu/docs/release/allwise/expsup/index.html}

\bibitem[{{Dark Energy Survey Collaboration} {et~al.}(2016){Dark Energy Survey
  Collaboration}, {Abbott}, {Abdalla}, {Aleksi{\'c}}, {Allam}, {Amara},
  {Bacon}, {Balbinot}, {Banerji}, {Bechtol}, {Benoit-L{\'e}vy}, {Bernstein},
  {Bertin}, {Blazek}, {Bonnett}, {Bridle}, {Brooks}, {Brunner}, {Buckley-Geer},
  {Burke}, {Caminha}, {Capozzi}, {Carlsen}, {Carnero-Rosell}, {Carollo},
  {Carrasco-Kind}, {Carretero}, {Castander}, {Clerkin}, {Collett}, {Conselice},
  {Crocce}, {Cunha}, {D'Andrea}, {da Costa}, {Davis}, {Desai}, {Diehl},
  {Dietrich}, {Dodelson}, {Doel}, {Drlica-Wagner}, {Estrada}, {Etherington},
  {Evrard}, {Fabbri}, {Finley}, {Flaugher}, {Foley}, {Fosalba}, {Frieman},
  {Garc{\'{\i}}a-Bellido}, {Gaztanaga}, {Gerdes}, {Giannantonio}, {Goldstein},
  {Gruen}, {Gruendl}, {Guarnieri}, {Gutierrez}, {Hartley}, {Honscheid}, {Jain},
  {James}, {Jeltema}, {Jouvel}, {Kessler}, {King}, {Kirk}, {Kron}, {Kuehn},
  {Kuropatkin}, {Lahav}, {Li}, {Lima}, {Lin}, {Maia}, {Makler}, {Manera},
  {Maraston}, {Marshall}, {Martini}, {McMahon}, {Melchior}, {Merson}, {Miller},
  {Miquel}, {Mohr}, {Morice-Atkinson}, {Naidoo}, {Neilsen}, {Nichol}, {Nord},
  {Ogando}, {Ostrovski}, {Palmese}, {Papadopoulos}, {Peiris}, {Peoples},
  {Percival}, {Plazas}, {Reed}, {Refregier}, {Romer}, {Roodman}, {Ross},
  {Rozo}, {Rykoff}, {Sadeh}, {Sako}, {S{\'a}nchez}, {Sanchez}, {Santiago},
  {Scarpine}, {Schubnell}, {Sevilla-Noarbe}, {Sheldon}, {Smith}, {Smith},
  {Soares-Santos}, {Sobreira}, {Soumagnac}, {Suchyta}, {Sullivan}, {Swanson},
  {Tarle}, {Thaler}, {Thomas}, {Thomas}, {Tucker}, {Vieira}, {Vikram},
  {Walker}, {Wechsler}, {Weller}, {Wester}, {Whiteway}, {Wilcox}, {Yanny},
  {Zhang}, \& {Zuntz}}]{DES}
{Dark Energy Survey Collaboration}, {Abbott}, T., {Abdalla}, F.~B., {et~al.}
  2016, \mnras, 460, 1270

\bibitem[{{Eisenstein} {et~al.}(2011){Eisenstein}, {Weinberg}, {Agol},
  {Aihara}, {Allende Prieto}, {Anderson}, {Arns}, {Aubourg}, {Bailey},
  {Balbinot}, \& et~al.}]{SDSS_main}
{Eisenstein}, D.~J., {Weinberg}, D.~H., {Agol}, E., {et~al.} 2011, \aj, 142, 72

\bibitem[{{Erfanianfar} {et~al.}(2013){Erfanianfar}, {Finoguenov}, {Tanaka},
  {Lerchster}, {Nandra}, {Laird}, {Connelly}, {Bielby}, {Mirkazemi}, {Faber},
  {Kocevski}, {Cooper}, {Newman}, {Jeltema}, {Coil}, {Brimioulle}, {Davis},
  {McCracken}, {Willmer}, {Gerke}, {Cappelluti}, \& {Gwyn}}]{Erfanianfar2013}
{Erfanianfar}, G., {Finoguenov}, A., {Tanaka}, M., {et~al.} 2013, The
  Astrophysical Journal, 765, 117

\bibitem[{{Garilli} {et~al.}(2014){Garilli}, {Guzzo}, {Scodeggio},
  {Bolzonella}, {Abbas}, {Adami}, {Arnouts}, {Bel}, {Bottini}, {Branchini},
  {Cappi}, {Coupon}, {Cucciati}, {Davidzon}, {De Lucia}, {de la Torre},
  {Franzetti}, {Fritz}, {Fumana}, {Granett}, {Ilbert}, {Iovino}, {Krywult}, {Le
  Brun}, {Le F{\`e}vre}, {Maccagni}, {Ma{\l}ek}, {Marulli}, {McCracken},
  {Paioro}, {Polletta}, {Pollo}, {Schlagenhaufer}, {Tasca}, {Tojeiro},
  {Vergani}, {Zamorani}, {Zanichelli}, {Burden}, {Di Porto}, {Marchetti},
  {Marinoni}, {Mellier}, {Moscardini}, {Nichol}, {Peacock}, {Percival},
  {Phleps}, \& {Wolk}}]{VIPERS}
{Garilli}, B., {Guzzo}, L., {Scodeggio}, M., {et~al.} 2014, \aap, 562, A23

\bibitem[{{Gerke} {et~al.}(2012){Gerke}, {Newman}, {Davis}, {Coil}, {Cooper},
  {Dutton}, {Faber}, {Guhathakurta}, {Konidaris}, {Koo}, {Lin}, {Noeske},
  {Phillips}, {Rosario}, {Weiner}, {Willmer}, \& {Yan}}]{Gerke}
{Gerke}, B.~F., {Newman}, J.~A., {Davis}, M., {et~al.} 2012, \apj, 751, 50

\bibitem[{{Hao} {et~al.}(2009){Hao}, {Koester}, {Mckay}, {Rykoff}, {Rozo},
  {Evrard}, {Annis}, {Becker}, {Busha}, {Gerdes}, {Johnston}, {Sheldon}, \&
  {Wechsler}}]{ecGMM}
{Hao}, J., {Koester}, B.~P., {Mckay}, T.~A., {et~al.} 2009, \apj, 702, 745

\bibitem[{{Kaiser} {et~al.}(2002){Kaiser}, {Aussel}, {Burke}, {Boesgaard},
  {Chambers}, {Chun}, {Heasley}, {Hodapp}, {Hunt}, {Jedicke}, {Jewitt},
  {Kudritzki}, {Luppino}, {Maberry}, {Magnier}, {Monet}, {Onaka}, {Pickles},
  {Rhoads}, {Simon}, {Szalay}, {Szapudi}, {Tholen}, {Tonry}, {Waterson}, \&
  {Wick}}]{PanSTARRS}
{Kaiser}, N., {Aussel}, H., {Burke}, B.~E., {et~al.} 2002, in \procspie, Vol.
  4836, Survey and Other Telescope Technologies and Discoveries, ed. J.~A.
  {Tyson} \& S.~{Wolff}, 154--164

\bibitem[{{Koester} {et~al.}(2007){Koester}, {McKay}, {Annis}, {Wechsler},
  {Evrard}, {Rozo}, {Bleem}, {Sheldon}, \& {Johnston}}]{Koester2007}
{Koester}, B.~P., {McKay}, T.~A., {Annis}, J., {et~al.} 2007, The Astrophysical
  Journal, 660, 221

\bibitem[{{Lawrence}(2013)}]{UKIDSS}
{Lawrence}, A. 2013, Astrophysics and Space Science Proceedings, 37, 271

\bibitem[{{Mehrtens} {et~al.}(2012){Mehrtens}, {Romer}, {Hilton},
  {Lloyd-Davies}, {Miller}, {Stanford}, {Hosmer}, {Hoyle}, {Collins}, {Liddle},
  {Viana}, {Nichol}, {Stott}, {Dubois}, {Kay}, {Sahl{\'e}n}, {Young}, {Short},
  {Christodoulou}, {Watson}, {Davidson}, {Harrison}, {Baruah}, {Smith},
  {Burke}, {Mayers}, {Deadman}, {Rooney}, {Edmondson}, {West}, {Campbell},
  {Edge}, {Mann}, {Sabirli}, {Wake}, {Benoist}, {da Costa}, {Maia}, \&
  {Ogando}}]{mehrtens+12}
{Mehrtens}, N., {Romer}, A.~K., {Hilton}, M., {et~al.} 2012, \mnras, 423, 1024

\bibitem[{{Merloni} {et~al.}(2012){Merloni}, {Predehl}, {Becker},
  {B{\"o}hringer}, {Boller}, {Brunner}, {Brusa}, {Dennerl}, {Freyberg},
  {Friedrich}, {Georgakakis}, {Haberl}, {Hasinger}, {Meidinger}, {Mohr},
  {Nandra}, {Rau}, {Reiprich}, {Robrade}, {Salvato}, {Santangelo}, {Sasaki},
  {Schwope}, {Wilms}, \& {German eROSITA Consortium}}]{merloni+12}
{Merloni}, A., {Predehl}, P., {Becker}, W., {et~al.} 2012, ArXiv e-prints
  [\eprint[arXiv]{1209.3114}]

\bibitem[{{Navarro} {et~al.}(1996){Navarro}, {Frenk}, \& {White}}]{NFW}
{Navarro}, J.~F., {Frenk}, C.~S., \& {White}, S.~D.~M. 1996, ApJ, 462, 563

\bibitem[{{Pineau}(2016)}]{Pineau2016}
{Pineau}, F., e.~a. 2016, {ARCHES XMatch cross-correlation tool}, in prep.

\bibitem[{{Pineau} {et~al.}(2015){Pineau}, {Boch}, {Derriere}, \& {Arches
  Consortium}}]{Pineau2015}
{Pineau}, F., {Boch}, T., {Derriere}, S., \& {Arches Consortium}. 2015, in
  Astronomical Society of the Pacific Conference Series, Vol. 495, Astronomical
  Society of the Pacific Conference Series, ed. A.~R. {Taylor} \&
  E.~{Rosolowsky}, 61

\bibitem[{{Planck Collaboration} {et~al.}(2014){Planck Collaboration}, {Ade},
  {Aghanim}, {Armitage-Caplan}, {Arnaud}, {Ashdown}, {Atrio-Barandela},
  {Aumont}, {Aussel}, {Baccigalupi}, \& et~al.}]{PlanckSZ}
{Planck Collaboration}, {Ade}, P.~A.~R., {Aghanim}, N., {et~al.} 2014, \aap,
  571, A29

\bibitem[{{Postman} {et~al.}(1996){Postman}, {Lubin}, {Gunn}, {Oke}, {Hoessel},
  {Schneider}, \& {Christensen}}]{postman}
{Postman}, M., {Lubin}, L.~M., {Gunn}, J.~E., {et~al.} 1996, \aj, 111, 615

\bibitem[{{Rosen} {et~al.}(2015){Rosen}, {Webb}, {Watson}, {Ballet}, {Barret},
  {Braito}, {Carrera}, {Ceballos}, {Coriat}, {Della Ceca}, {Denkinson},
  {Esquej}, {Farrell}, {Freyberg}, {Gris{\'e}}, {Guillout}, {Heil},
  {Law-Green}, {Lamer}, {Lin}, {Martino}, {Michel}, {Motch}, {Nebot
  Gomez-Moran}, {Page}, {Page}, {Page}, {Pakull}, {Pye}, {Read}, {Rodriguez},
  {Sakano}, {Saxton}, {Schwope}, {Scott}, {Sturm}, {Traulsen}, {Yershov}, \&
  {Zolotukhin}}]{3XMMDR5}
{Rosen}, S.~R., {Webb}, N.~A., {Watson}, M.~G., {et~al.} 2015, ArXiv e-prints
  [\eprint[arXiv]{1504.07051}]

\bibitem[{{Rykoff} {et~al.}(2014){Rykoff}, {Rozo}, {Busha}, {Cunha},
  {Finoguenov}, {Evrard}, {Hao}, {Koester}, {Leauthaud}, {Nord}, {Pierre},
  {Reddick}, {Sadibekova}, {Sheldon}, \& {Wechsler}}]{Rykoff2013}
{Rykoff}, E.~S., {Rozo}, E., {Busha}, M.~T., {et~al.} 2014, The Astrophysical
  Journal, 785, 104

\bibitem[{{Shanks} {et~al.}(2015){Shanks}, {Metcalfe}, {Chehade}, {Findlay},
  {Irwin}, {Gonzalez-Solares}, {Lewis}, {Yoldas}, {Mann}, {Read}, {Sutorius},
  \& {Voutsinas}}]{ATLAS}
{Shanks}, T., {Metcalfe}, N., {Chehade}, B., {et~al.} 2015, \mnras, 451, 4238

\bibitem[{{Takey} {et~al.}(2013){Takey}, {Schwope}, \& {Lamer}}]{Takey2}
{Takey}, A., {Schwope}, A., \& {Lamer}, G. 2013, \aap, 558, A75

\bibitem[{{Takey} {et~al.}(2014){Takey}, {Schwope}, \& {Lamer}}]{Takey3}
{Takey}, A., {Schwope}, A., \& {Lamer}, G. 2014, \aap, 564, A54

\bibitem[{{Wen} \& {Han}(2011)}]{Wen2011}
{Wen}, Z.~L. \& {Han}, J.~L. 2011, The Astrophysical Journal, 734, 68

\bibitem[{Wen {et~al.}(2009)Wen, Han, \& Liu}]{Wen2009}
Wen, Z.~L., Han, J.~L., \& Liu, F.~S. 2009, The Astrophysical Journal
  Supplement Series, 183, 197

\bibitem[{{Zwicky} {et~al.}(1961){Zwicky}, {Herzog}, {Wild}, {Karpowicz}, \&
  {Kowal}}]{Abell}
{Zwicky}, F., {Herzog}, E., {Wild}, P., {Karpowicz}, M., \& {Kowal}, C.~T.
  1961, {Catalogue of galaxies and of clusters of galaxies, Vol. I}

\end{thebibliography}

\appendix
\section{Definition of the $p_\nu(\chi^2)$ function} \label{sec:pnu}

Here the definition of one of the key functions for the ICF is given.
We start from the mean colour values $\cavg$ and the covariance matrix $\covariance$ 
at redshift $z$. If there are $n$ colours, the dimensions of $\cavg,$ and $\covariance$  are $n$ and $n \times n$, respectively.

We consider a galaxy with magnitudes $f_{1..m}$ and errors $\delta f_{1..m}$  measured in $m$ frequency bands.
It is not necessary that all $m$ magnitudes are measured for each galaxy in the catalogue, i.e. some flux values might be undefined. 

We also define a set of index pairs $I_i = (I_{1,i}, I_{2,i})$, where $I_{1,i},
I_{2,i}$ are two band indices and each pair defines a colour.

For example, we take SDSS-UKIDSS-AllWISE band set (u,g,r,i,z,Y,J,K,W1,W2), 
then $m=10$.
We then define $n=9$ colours: u-g,g-r,r-i,i-z,r-Y,Y-K,Y-W1,r-W1,W1-W2. 
The J band was dropped from the list of colours, as J-K changes only little over the 
desired redshift range. Thus:
I = [(1,2), (2,3), (3,4), (4,5), (3,6), (6,8), (8,9), (3,9), (9,10)]

Then for a galaxy the colour offset 
$c_i = f_{I_{1,i}} - f_{I_{2,i}} - \cavg_i$ is calculated with 
$i = 1..n$, and colour errors $\delta c_i^2 = \delta f_{I_{1,i}}^2+ \delta f_{I_{2,i}}^2$.

The error covariance matrix $\mathbf{C}$ defined as
\begin{equation}
 \mathbf{C}_{pq} = \left\{ 
  \begin{array}{l l}
    \delta c_{p}^2, & p = q \\
    \delta f_{I_{1,p}}^2, & p \neq q\ and\ (I_{1,p} = I_{1,q}\ or\ I_{2,p} = I_{2,q}) \\
    -\delta f_{I_{1,p}}^2, & p \neq q\ and\ (I_{1,p} = I_{2,q}\ or\ I_{2,p} = I_{1,q}) \\
    0, & {\rm otherwise}
  \end{array}\right.
.\end{equation}

Examples of $\mathbf{C}_{pq}$ matrix elements are as follows:
\begin{enumerate}
 \item $C_{23}$, i.e. $g-r$ versus~$r-i$ colour correlation coefficient, has $r$ as the 
 second band in $g-r$ and as the first band in $r-i$, thus $I_{2,p} = \mlq r \mrq = 3 = I_{1,q}$ and $C_{23} = -\delta f_{r}^2$.
 \item $C_{39}$, i.e. $r-i$ versus~$r-W1$ colour, has $r$ as the first band in both colours, 
thus $I_{1,p} = \mlq r \mrq = 3 = I_{1,q}$ and $C_{39} = \delta f_{r}^2$.
\item $C_{33}$, i.e.~simple r-i colour error, $C_{33} = \delta c_{r-i}^2$.
\end{enumerate}

For each redshift $z$ and one has to calculate
\begin{equation}
 \chi^2(z) = \mathbf{c} (\covariance + \mathbf{C})^{-1} \mathbf{c} \label{eq:chi}
.\end{equation} 

If one or several bands are missing for a given source, then the 
corresponding elements of $\mathbf{c}$ and rows/columns from $\covariance$ 
and $\mathbf{C}$ are removed.

If one of the colours is the exact combination of two or more other colours then the covariance
matrix defined above is zero. In this case, one of the three colours should be removed. For example,
for SDSS-UKIDSS-AllWISE, $c_{r-Y} + c_{Y-W1} = c_{r-W1}$. If the r, Y, and W1 magnitudes are known for a given galaxy, then only $c_{r-Y}, c_{Y-W1}$ should be used. If one of the r, Y, and W1 magnitudes is not known then we are left with only one colour.

The value $\chi_j^2(z)$ in Eq.~\ref{eq:chi} is assumed to follow a 
$\chi^2$-distribution. 
Although colours and colour errors are obviously non-independent variables,
it is assumed they would be independent and this issue is neglected here.

\end{document}